\documentclass[acmsmall,screen]{acmart}

\usepackage{graphicx}
\usepackage{textcomp}
\usepackage{multirow}
\usepackage{longtable}
\usepackage{lscape}
\usepackage{color, colortbl}
\usepackage{rotating}
\usepackage{wrapfig}
\usepackage{subcaption}
\usepackage{pdflscape}
\usepackage{hyperref}
\usepackage{array}
\usepackage{pifont}
\usepackage{balance}
\usepackage[most]{tcolorbox}
\usepackage{tikz}
\usepackage[normalem]{ulem}
\usepackage{url}
\usepackage{soul}
\usepackage{threeparttable}
\usepackage{xspace}
\usepackage{diagbox}
\usepackage{booktabs}
\usepackage{adjustbox}
\usepackage{svg}
\usepackage[ruled,vlined,lined,commentsnumbered]{algorithm2e}
\usepackage{enumitem}
\usepackage{fancybox}
\usepackage{fontenc}
\usepackage{listings}
\usepackage{longtable}
\usepackage{lscape}
\usepackage{makecell}
\usepackage{marvosym}
\usepackage{multicol}
\usepackage{multirow}
\usepackage{pifont}%
\usepackage{rotating}
\usepackage{setspace}
\usepackage[most]{tcolorbox}
\usepackage{threeparttable}
\usepackage{tikz}
\usepackage[normalem]{ulem}
\usepackage{url}
\usepackage{soul}
\usepackage{wasysym}
\usepackage{textcomp}
\usepackage{xcolor}
\usepackage{wrapfig}
\usepackage{pdflscape}

\newcommand{\intuition}[1]{
\begin{tcolorbox}[colback=white,boxrule=1pt,top=0pt,bottom=0pt,left=1pt,right=2pt,top=2pt,bottom=2pt]
\em #1
\end{tcolorbox}
}

\begin{document}
\newcommand{\ModelName}{SolAgent}

\newcommand{\toolname}{{\sc SolAgent}\xspace}
\newcommand{\toolnameb}{{\bf SolAgent}\xspace}

\title{{\ModelName:} A Specialized Multi-Agent Framework for Solidity Code Generation}

\author{Wei Chen}
\affiliation{
  \institution{Shanghai Jiao Tong University}
  \city{Shanghai}
  \country{China}
  }
\email{chenwei8@sjtu.edu.cn}

\author{Zhiyuan Peng}
\affiliation{
  \institution{Shanghai Jiao Tong University}
  \city{Shanghai}
  \country{China}
  }
\email{pzy2000@sjtu.edu.cn}

\author{Xin Yin}
\affiliation{
  \institution{The State Key Laboratory of Blockchain and Data Security, Zhejiang University}
  \city{Hangzhou}
  \country{China}
  }
\email{xyin@zju.edu.cn}

\author{Chao Ni}
\affiliation{
  \institution{The State Key Laboratory of Blockchain and Data Security, Zhejiang University}
  \city{Hangzhou}
  \country{China}
  }
\email{chaoni@zju.edu.cn}

\author{Chenhao Ying}
\authornote{Corresponding authors.}
\affiliation{
  \institution{Shanghai Jiao Tong University}
  \city{Shanghai}
  \country{China}
  }
\email{yingchenhao@sjtu.edu.cn}

\author{Bang Xie}
\affiliation{
  \institution{Shanghai Jiao Tong University}
  \city{Shanghai}
  \country{China}
  }
\email{xiebang-1213@sjtu.edu.cn}

\author{Yuan Luo}
\authornotemark[1]
\affiliation{
  \institution{Shanghai Jiao Tong University}
  \city{Shanghai}
  \country{China}
  }
\email{yuanluo@sjtu.edu.cn}

\begin{abstract}
Smart contracts are the backbone of the decentralized web, yet ensuring their functional correctness and security remains a critical challenge. While Large Language Models (LLMs) have shown promise in code generation, they often struggle with the rigorous requirements of smart contracts, frequently producing code that is buggy or vulnerable. To address this, we propose \toolname, a novel tool-augmented multi-agent framework that mimics the workflow of human experts. \toolname integrates a \textbf{dual-loop refinement mechanism}: an inner loop using the \textit{Forge} compiler to ensure functional correctness, and an outer loop leveraging the \textit{Slither} static analyzer to eliminate security vulnerabilities. Additionally, the agent is equipped with file system capabilities to resolve complex project dependencies. Experiments on the SolEval+ Benchmark, a rigorous suite derived from high-quality real-world projects, demonstrate that \toolname achieves a Pass@1 rate of up to \textbf{64.39\%}, significantly outperforming state-of-the-art LLMs ($\sim$25\%), AI IDEs (e.g., GitHub Copilot), and existing agent frameworks. Moreover, it reduces security vulnerabilities by up to \textbf{39.77\%} compared to human-written baselines. Finally, we demonstrate that the high-quality trajectories generated by \toolname can be used to distill smaller, open-source models, democratizing access to secure smart contract generation. We release our data and code at \url{https://github.com/openpaperz/SolAgent}.
\end{abstract}


\maketitle

\section{Introduction}
\label{sec:introduction}

Smart contracts, self-executing programs running on blockchains like Ethereum, have become the cornerstone of the decentralized web (Web3), managing billions of dollars in digital assets. Unlike traditional software, smart contracts are immutable once deployed; any vulnerability can lead to catastrophic financial losses~\cite{Chen2017Underoptimized,Chen2024Angels}. The critical nature of smart contract security was demonstrated in May 2025 when Cetus Protocol suffered a devastating exploit that drained over \$260 million through manipulated price curves and flawed reserve calculations~\cite{BinanceNews2025Cetus}. This incident, along with high-profile historical attacks and recent DeFi exploits~\cite{Zhang2023Demystifying,Perez2021Smart,Sharma2023Mixedmethods}, underscores the persistent vulnerability of smart contract systems. Recent empirical studies have revealed the severity and prevalence of smart contract vulnerabilities~\cite{Zhang2023Demystifying,Wang2025Unity}. Furthermore, systematic reviews~\cite{Azimi2025Systematic} have shown that existing security design patterns can only directly mitigate a limited subset of known vulnerability types, indicating significant gaps in current defensive practices. Consequently, ensuring the functional correctness and security of smart contract code is of paramount importance.

Recent advancements in Large Language Models (LLMs) have demonstrated impressive capabilities in automated code generation~\cite{Yin2024ThinkRepair,Liu2024Make,Liao2024MathbfA^3A3CodGen}. Models like GPT-5 and Claude-Sonnet-4.5 have shown proficiency in generating code snippets across various languages. However, when applied to smart contract development, these general-purpose models often fall short. They frequently produce code that is syntactically plausible but functionally incorrect (``hallucinations'') or, worse, contains subtle security vulnerabilities that are difficult to detect manually~\cite{Sun2024GPTScan,Wu2024AdvSCanner,Wang2024SmartInv}. Furthermore, existing LLM-based agent frameworks~\cite{Hong2024MetaGPT,Qian2024ChatDev,Zhang2024AutoCodeRover} typically treat code generation as a text-processing task, lacking integration with domain-specific verification tools. They often struggle with the complex dependency structures of real-world Solidity projects, leading to code with compilation errors or runtime failures in contract interactions.

To address these challenges, we propose \toolname, a novel tool-augmented multi-agent framework designed specifically for high-quality smart contract generation. Unlike ``black-box'' generation approaches, \toolname mimics the workflow of human experts by integrating a \textbf{dual-loop refinement mechanism}. In the inner loop, the agent utilizes the \textit{Forge}~\cite{FoundryContributors2025Foundry} compiler and testing framework to iteratively verify and fix functional errors, ensuring the code meets the specified requirements. In the outer loop, it leverages the \textit{Slither}~\cite{Feist2019Slither,TrailofBits2025Slither} static analyzer to detect and remediate security vulnerabilities. Additionally, \toolname is equipped with file system tools, enabling it to explore project structures and resolve dependencies contextually, a capability often limited in general-purpose agents. 
To investigate the effectiveness of \toolname, we conduct experiments on the \textbf{SolEval+ Benchmark}, a rigorous suite constructed from real-world high-quality smart contracts. Our experiments demonstrate that \toolname significantly outperforms state-of-the-art LLMs, AI-powered IDEs (e.g., GitHub Copilot), and existing agent frameworks. Specifically, it achieves a Pass@1 rate of \textbf{64.39\%} (compared to $\sim$25\% for vanilla LLMs) and reduces security vulnerabilities by nearly \textbf{39.77\%} compared to human-written baselines. Furthermore, we show that the high-quality interaction trajectories generated by \toolname can be used to distill smaller, open-source models (e.g., Qwen3-8B), enabling efficient and accessible smart contract generation.

In summary, this paper makes the following contributions:

\textbf{A. Tool-Augmented Multi-Agent Framework:} We propose \toolname, the first framework that integrates domain-specific compilation (Forge) and static analysis (Slither) tools for end-to-end smart contract generation and refinement.

\textbf{B. Dual-Loop Refinement Mechanism:} We introduce a dual-loop mechanism that simultaneously optimizes for functional correctness and security, effectively addressing the ``impossible triangle'' of single-pass generation.

\textbf{C. Extensive Evaluations:} We conduct extensive evaluations showing that \toolname establishes a new state-of-the-art in automated smart contract generation, and demonstrate the effectiveness of using agent trajectories for model distillation.

\section{Background and Motivation}
\label{sec:motivation}

Smart contract development demands a rigorous balance between \textit{functional correctness} and \textit{security assurance}. 
Given the immutable nature of blockchains and the substantial financial assets involved, deployed code must not only execute the intended logic flawlessly but also withstand adversarial attacks.
In this section, we aim to explore the limitations of existing \textit{LLM-based code generation} and \textit{general-purpose multi-agent frameworks} in the context of Solidity development.

\subsection{Executability and Security Bottlenecks in LLM-based Generation}

While Large Language Models like GPT and Claude have demonstrated impressive coding capabilities, they face a dual challenge when applied to Solidity: \textit{low executability} and \textit{latent vulnerability}.
The primary barrier to automated smart contract generation is the correctness bottleneck. General LLMs often prioritize syntactical plausibility over strict Solidity compliance, leading to hallucinations of non-existent libraries or syntax errors that cause compilation failures. As a result, a significant portion of generated code fails to execute or pass basic unit tests.
Even when the code compiles, LLMs lack the intrinsic awareness of functional correctness and security patterns. They tend to generate code that remains vulnerable to edge cases (e.g., reentrancy, integer overflow)~\cite{Sun2024GPTScan,Wu2024AdvSCanner,Wang2024SmartInv}. These issues arise because LLMs operate in a ``single-pass'' manner, lacking the iterative refinement and domain-specific validation required for production-grade smart contracts.

To verify this, we conducted a preliminary study, as shown in Table~\ref{tab:motivation}.
We prompted GPT-5.1 and Claude-Sonnet-4.5 to generate 50 smart contracts based on requirements from a standard dataset.
We measured the \textit{Compilation Failure Rate} and then analyzed the contracts using unit tests (for functional correctness) and Slither (for security).
Results show that while top-tier models can generate text, a large portion fails to compile. Among those that compiled, a significant percentage failed to pass unit tests or contained low-to-high severity vulnerabilities.
This indicates that relying solely on general LLMs is insufficient for engineering reliable smart contracts.

\begin{table}[htbp]
\vspace{-0.1cm}
\centering
\caption{Performance Gaps in General LLMs (Preliminary Study)}
\vspace{-0.3cm}
\resizebox{.55\linewidth}{!}{%
	\begin{tabular}{lccc}
		\toprule
		\textbf{Models} & \makecell{\textbf{Compilation}\\\textbf{Failure (\%)}} & \makecell{\textbf{Test}\\\textbf{Failure (\%)}} & \makecell{\textbf{Vulnerable}\\\textbf{Contracts (\%)}} \\
		\midrule
		GPT-5.1 & 54.0\% & 87.02\% & 20.0\% \\
		Claude-Sonnet-4.5 & 58.0\% & 82.60\% & 14.0\% \\
		GPT-5-Mini & 60.0\% & 82.32\% & 14.0\% \\
		\bottomrule
	\end{tabular}%
}
\label{tab:motivation}
\vspace{-.3cm}
\end{table} 

\subsection{Domain Mismatch in General Multi-Agent Frameworks}

Multi-agent frameworks, such as MetaGPT~\cite{Hong2024MetaGPT} and ChatDev~\cite{Qian2024ChatDev}, have emerged to handle complex software tasks by simulating software development teams.
However, these frameworks suffer from a \textit{domain mismatch} when applied to the rigorous environment of Solidity.
They struggle to generate complex, functionally correct code while simultaneously maintaining security standards, often sacrificing one for the other.

Existing frameworks typically adopt a standard ``Waterfall'' workflow (Requirements $\to$ Design $\to$ Coding $\to$ Testing) designed for general-purpose software (e.g., Python, Java web apps).
They lack the verification-driven feedback loops essential for smart contracts.
For instance, a general ``Tester'' agent might generate simple input/output tests but fails to utilize Solidity-specific tools (e.g., Forge, Slither) to detect logic errors or security vulnerabilities.
Consequently, when encountering compilation errors or logical bugs, these general agents often lack the specialized domain knowledge required for remediation, resulting in a degraded feedback loop of ineffective modifications~\cite{Huang2024Large}.

\intuition{{\bf Intuition.}
To address these limitations, a specialized framework is needed.
We posit that high-quality Solidity generation requires a Specialized Dual-Loop Mechanism: 
(1) a Correctness Loop driven by compiler feedback and unit tests to ensure high executability (Pass@1), and 
(2) a Security Loop driven by static analysis to constrain the solution within safe boundaries.
By incorporating specialized roles (e.g., Auditors) and integrating domain tools directly into the workflow, we can achieve high functional correctness without compromising security.
}

\section{Approach}

\begin{figure*}[htbp]
    \vspace{-0.3cm}
    \centering
    \includegraphics[width=.83\linewidth]{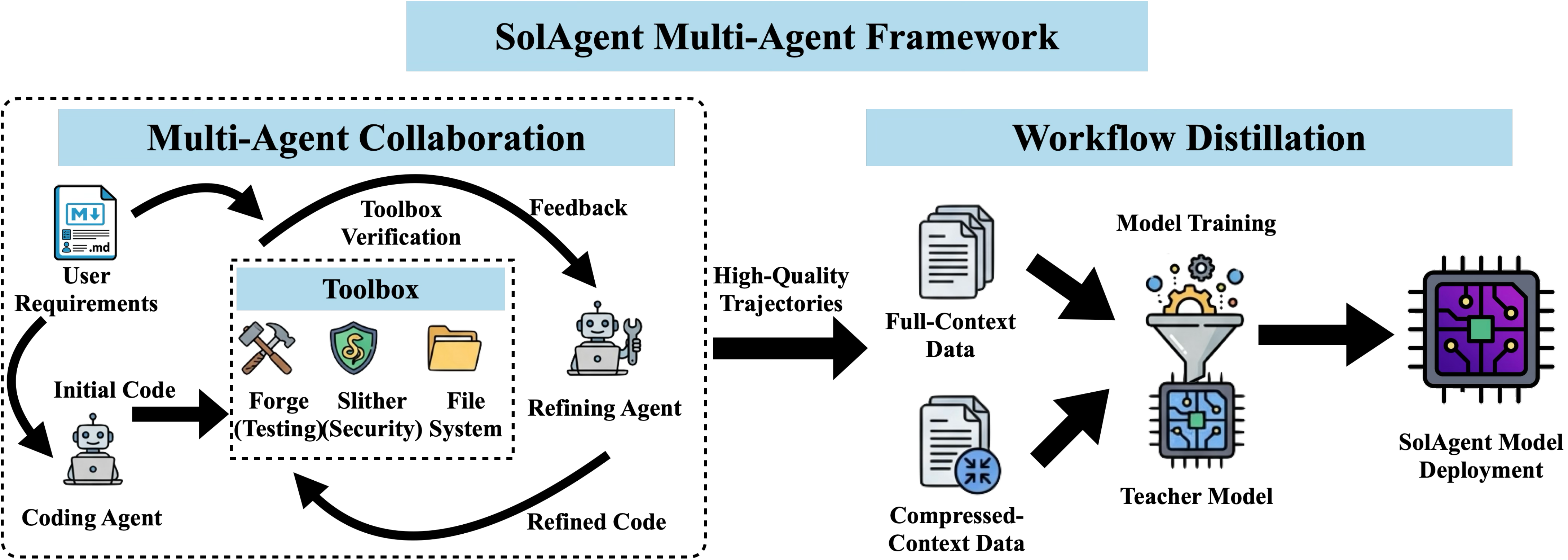}
    \caption{Overview of \toolname}
    \vspace{-0.5cm}
    \label{fig:overview}
\end{figure*}

\subsection{Overview}
We propose \toolname, a multi-agent framework for secure and gas-efficient smart contract generation. 
It addresses the limitations of single-pass LLM generation by integrating iterative refinement with external verification tools.
As illustrated in Fig.~\ref{fig:overview}, the workflow is as follows:
\begin{itemize}[leftmargin=*]
    \item \textbf{Multi-Agent Collaboration}: We employ a dual-agent system where a \textit{Coding Agent} generates initial solutions and a \textit{Refining Agent} iteratively improves them based on feedback.
    \item \textbf{Tool-Augmented Refinement}: We leverage industry-standard tools, \textit{Forge} and \textit{Slither}, to provide rigorous feedback on correctness, security, and gas efficiency. Additionally, the agents are equipped with file system tools to explore and understand the codebase context.
    \item \textbf{Workflow Distillation}: We distill the high-quality trajectories generated by the agent into a lightweight open-source model (Qwen3-8B) to enable efficient, one-shot generation.
\end{itemize}

\subsection{Module 1: Multi-Agent Collaboration Framework}
To achieve high-quality code generation, we adopt a multi-agent architecture that mimics the human software development lifecycle. This module involves two key roles:

\begin{itemize}[leftmargin=*]
    \item \textbf{Coding Agent}: Acts as the primary developer. It analyzes the requirements and generates the initial Solidity smart contract implementation.
    \item \textbf{Refining Agent}: Acts as the reviewer and optimizer. It analyzes the execution feedback and security reports, and directly modifies the code to fix bugs and optimize gas usage.
\end{itemize}

\toolname interact in a loop. 
In each round $t$, the Refining Agent receives feedback $F_t$ from the environment and generates refined code $C_{t+1}$. 
This process continues until the stopping criteria are met, as detailed in Section~\ref{sec:exp-setting}.

\subsection{Module 2: Tool-Augmented Refinement}
Large Language Models often struggle with subtle logic bugs and security vulnerabilities. In this section, we aim to harness external symbolic execution and static analysis tools to ground the LLM's reasoning. To this end, we address three key tasks:

\subsubsection{Task 1: Correctness and Efficiency Verification via Forge}
To ensure functional correctness and gas efficiency, we integrate \textit{Forge}, a fast and flexible Ethereum testing framework.
For each generated contract $C_t$, we execute the accompanying test suite $T$.
The feedback $F_{forge}$ includes:
\begin{itemize}[leftmargin=*]
    \item \textbf{Pass Rate}: The ratio of passed tests to total tests.
    \item \textbf{Failure Details}: Specific assertion failures and stack traces for failed tests, allowing the agent to pinpoint logical errors.
\end{itemize}

\subsubsection{Task 2: Security Analysis via Slither}
Beyond correctness, smart contracts must be secure. We employ \textit{Slither}, a static analysis framework, to detect vulnerabilities.
\begin{itemize}[leftmargin=*]
    \item \textbf{Vulnerability Detection}: Slither scans $C_t$ for known vulnerability patterns. We categorize alerts into High, Medium, and Low severity. The feedback $F_{slither}$ guides the agent to fix critical security flaws.
\end{itemize}

\subsubsection{Task 3: Context Exploration via File System Tools}
To ensure the generated code integrates seamlessly with the existing project structure, the Refining Agent is equipped with file system tools:
\begin{itemize}[leftmargin=*]
    \item \textbf{List Directory}: Allows the agent to inspect the project structure and locate relevant files.
    \item \textbf{Read File}: Enables the agent to read dependency files, interfaces, and utility libraries to ensure correct imports and usage.
\end{itemize}

\subsubsection{Task 4: Dynamic Stopping Mechanism}
Standard iterative refinement often suffers from infinite loops or degradation. To address this, we propose a \textit{Dynamic Stopping Algorithm}.
The intuition is that refinement should cease when the code converges (no changes), oscillates (loops), or achieves all objectives.
Formally, we define the stopping condition based on three factors:
\begin{enumerate}[leftmargin=*]
    \item \textbf{Success}: All tests pass and no critical vulnerabilities remain.
    \item \textbf{Stagnation}: The pass rate does not improve for $N$ consecutive rounds.
    \item \textbf{Oscillation}: The feedback similarity between consecutive rounds exceeds a threshold $\tau$.
\end{enumerate}

We calculate the feedback similarity $Sim(F_t, F_{t-1})$ using the sequence matching ratio. If $Sim(F_t, F_{t-1}) > \tau$, it indicates the agent is stuck in a loop receiving identical errors (e.g., persistent compilation errors), and we terminate the process to save resources.

\begin{algorithm}[!ht]
\small
\caption{Tool-Augmented Refinement Loop}\label{alg:refinement}
\LinesNumbered
\KwIn{Initial Code $C_0$, Test Suite $T$, Max Rounds $MaxRounds$}
\KwOut{Best Refined Code $C_{best}$}

$C_{best} \leftarrow C_0$, $Score_{best} \leftarrow 0$\;

\For{$t \leftarrow 1$ \KwTo $MaxRounds$}{
    $F_{forge} \leftarrow \text{RunForge}(C_{t-1}, T)$\;
    $F_{slither} \leftarrow \text{RunSlither}(C_{t-1})$\;
    $F_t \leftarrow \text{Aggregate}(F_{forge}, F_{slither})$\;
    
    \If{$\text{IsPerfect}(F_t) \lor \text{IsStagnant}(F_t, N) \lor \text{IsLooping}(F_t, F_{t-1})$}{
        \textbf{break}\;
    }
    
    $C_t \leftarrow \text{RefineAgent}(C_{t-1}, F_t)$\;
    
    \If{$\text{Score}(C_t) > Score_{best}$}{
        $C_{best} \leftarrow C_t$\;
    }
}
\Return{$C_{best}$}
\end{algorithm}

\subsection{Module 3: Workflow Distillation}
While large language models have shown promising performance in multi-agent systems, deploying such systems requires substantial computational resources, making them prohibitively expensive for small organizations and individual users. 
To transfer these capabilities to smaller models and reduce deployment and usage costs, we perform \textit{Workflow Distillation}.

\subsubsection{Data Collection}
We collect interaction trajectories from the multi-agent system.
We distinguish between two types of data:
\begin{itemize}[leftmargin=*]
    \item \textbf{Full-Context Trajectories}: The agent generates code based on the original requirements $R_{full}$, which contain detailed, human-labeled comments. This provides rich semantic information but consumes more tokens.
    \item \textbf{Compressed-Context Trajectories}: We employ a summarization step to compress the verbose comments in $R_{full}$ into concise descriptions $R_{summary}$. The agent generates code based on this token-efficient input. This dual-source data strategy ensures the distilled model is robust to varying input verbosity and efficient with context usage.
\end{itemize}

\subsubsection{Model Training}
We filter the trajectories to retain only those that resulted in fully verified and secure contracts.
We then fine-tune a \textit{Qwen3-8B} model on this distilled dataset.
The objective is to minimize the negative log-likelihood of the optimal code $C^*$ given the requirements $R$:
\begin{equation}
    \mathcal{L} = -\sum_{i} \log P(C^*_i | R, C^*_{<i})
\end{equation}
This process effectively ``compresses'' the multi-agent intelligence into a single, efficient model, enabling it to produce secure and correct code in a single pass.

\section{Experiment}
\label{sec:experiment}

In this section, we first describe the datasets used for evaluation and fine-tuning. Then, we introduce the baseline approaches and the evaluation metrics. Finally, we provide the implementation details and experimental settings.

\begin{figure*}[htbp]
    \vspace{-0.2cm}
    \centering
    \includegraphics[width=.96\linewidth]{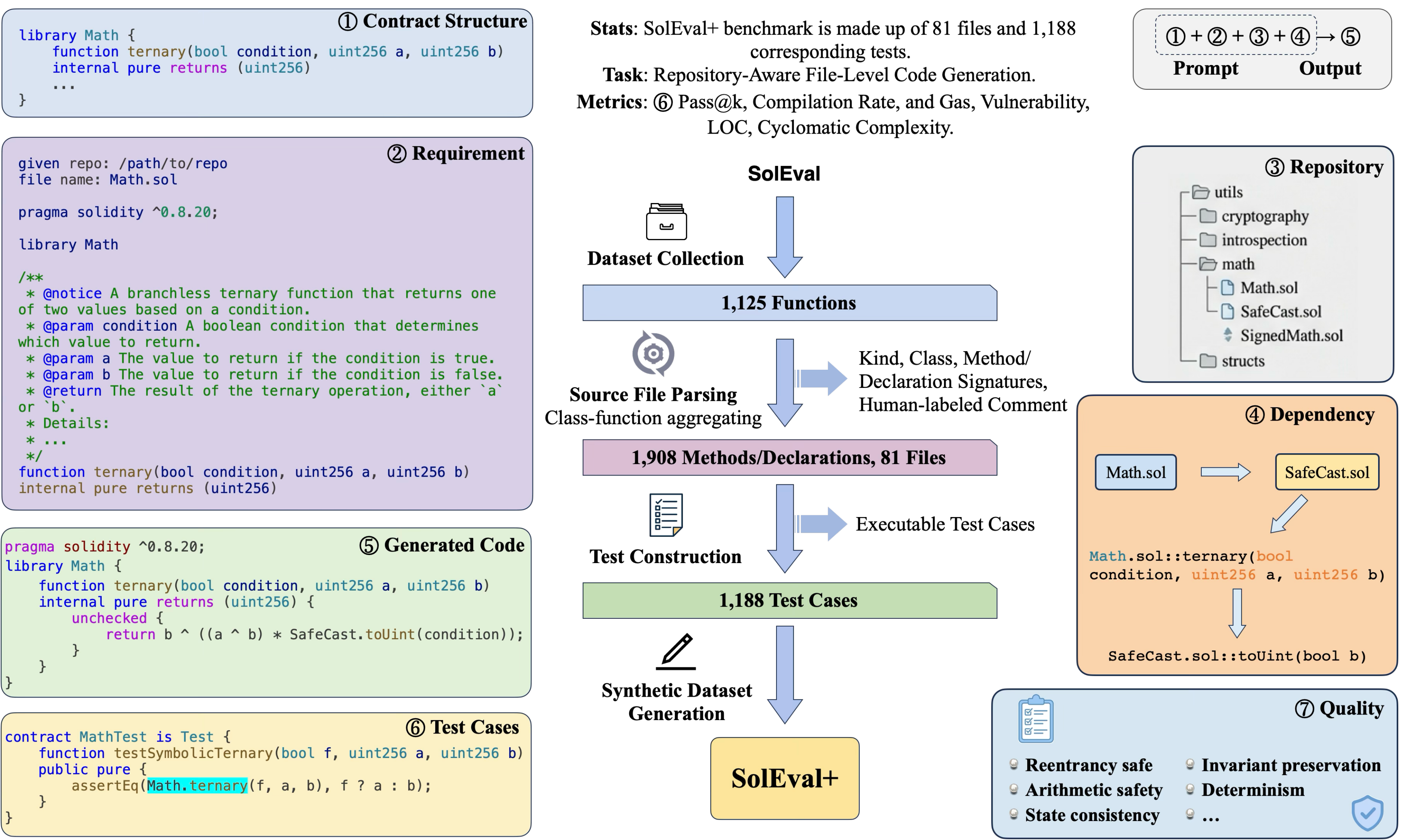}
    \caption{The SolEval+ Benchmark}
    \vspace{-0.5cm}
    \label{fig:soleval-plus-combine}
\end{figure*}

\subsection{Datasets}
We evaluate on the SolEval+ Benchmark, constructed from the SolEval~\cite{Peng2025SolEval}, and fine-tune our distilled model on a high-quality trajectory dataset generated by our multi-agent system.

\subsubsection{SolEval+ Benchmark}
To rigorously evaluate file-level smart contract generation with comprehensive testing and security validation, we constructed the SolEval+ Benchmark (Figure~\ref{fig:soleval-plus-combine}) by extending the repository-aware function-level SolEval dataset~\cite{Peng2025SolEval}.
We augmented it with executable unit tests and security evaluation infrastructure through a systematic three-step construction pipeline, as illustrated in the figure.

\textbf{Dataset Collection:} We collected 1,125 functions from the SolEval dataset, which originally contains repository-level Solidity projects with natural language specifications.

\textbf{Source File Parsing:} We performed class-function aggregation, extracting structured metadata including contract kind (e.g., library, interface, contract), class names, method/declaration signatures, and human-labeled comments that describe each function's intended behavior.

\textbf{Test Construction:} We constructed 1,188 executable test cases using the Foundry framework, covering critical functionalities such as state transitions, edge cases, access control, and event emissions, with each test manually reviewed for correctness.
Through this pipeline, we transformed the original function-level specifications into a file-level benchmark with comprehensive test coverage, enabling automated evaluation of functional correctness (Pass@k), security vulnerabilities (Vulnerability Count), and gas efficiency.


\subsubsection{Fine-tuning Dataset}
\label{sec:ft-dataset}
To distill the capabilities of our multi-agent system into a smaller model, we collected interaction trajectories from \toolname running on the training split of the SolEval+ Benchmark.
We processed the data as follows:
\begin{itemize}[leftmargin=*]
    \item \textbf{Data Collection}: We recorded the dialogue history of the Coding Agent (1 round) and Refining Agent ($\ge 1$ rounds). Each conversation round is treated as a training sample.
    \item \textbf{Cleaning \& Normalization}: We retained \texttt{system}, \texttt{user}, \texttt{assistant}, and \texttt{tool} messages, preserving \texttt{tool\_calls} and content. Irrelevant metadata was removed to unify the format.
    \item \textbf{Dataset Construction}: We created two variants:
    \begin{itemize}
        \item \texttt{dataset\_tracker}: Contains raw user queries without summaries.
        \item \texttt{dataset\_mix}: A mixture of trajectories from \texttt{dataset\_tracker} and new trajectories generated using summarized user queries. In the latter, the original detailed requirements in SolEval+ are replaced with concise functional summaries to prompt the agents.
    \end{itemize}
    \item \textbf{Data Split}: We randomly split the constructed datasets into a training set (80\%) and a held-out test set (20\%) to avoid data leakage. The interaction trajectories were collected strictly from the training set, while the final evaluation for distilled models was conducted on the test set.
\end{itemize}

\begin{figure*}[htbp]
    \vspace{-0.1cm}
    \centering
    \includegraphics[width=.98\linewidth]{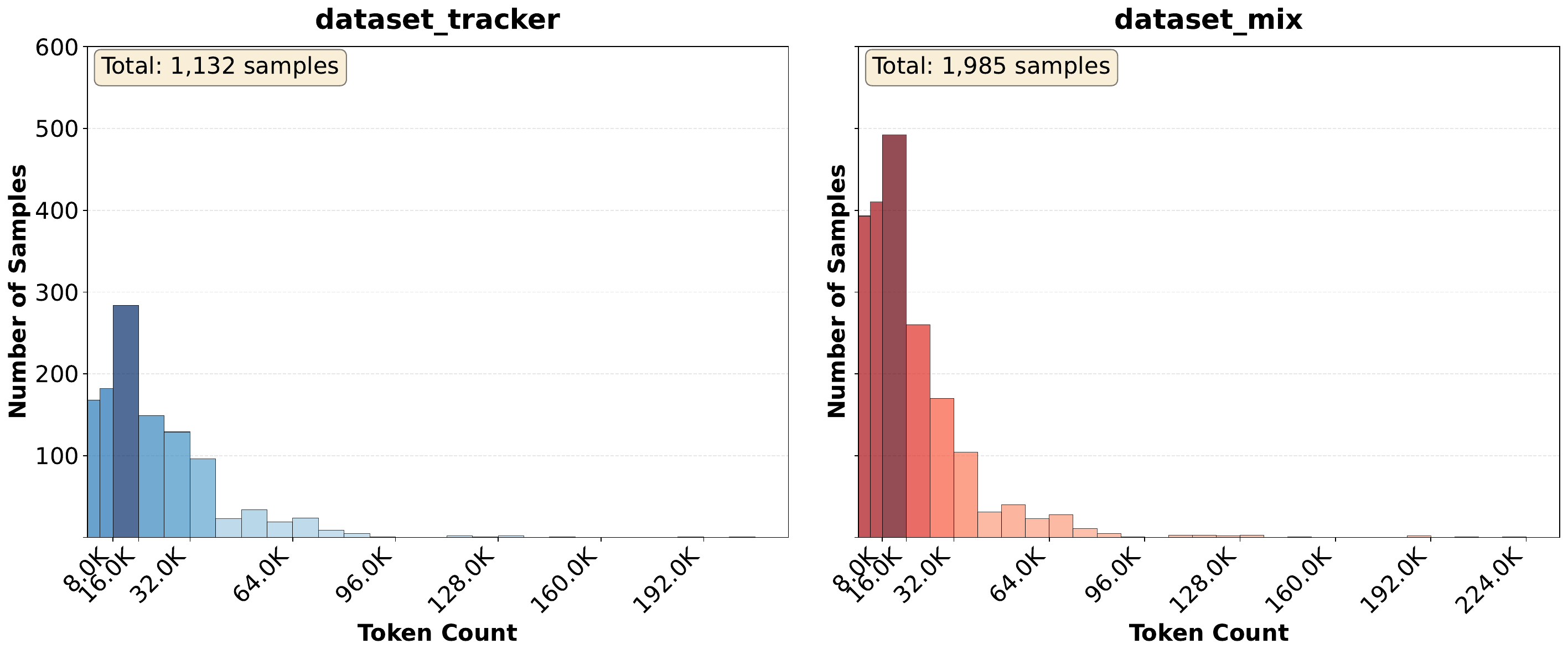}
    \vspace{-0.2cm}
    \caption{Dataset Token Distribution}
    \vspace{-0.5cm}
    \label{fig:token-distribution-comparison}
\end{figure*}

The statistics of the fine-tuning datasets are presented in Table~\ref{tab:dataset}.
Figure~\ref{fig:token-distribution-comparison} shows the distribution of token counts per sample in the two fine-tuning datasets.
In both datasets, most samples fall in the 0--32K token range (e.g., \texttt{dataset\_tracker}: 350 in [0, 8K) and 284 in [8K, 16K); \texttt{dataset\_mix}: 803 in [0, 8K) and 492 in [8K, 16K)), with a long tail extending beyond 64K.

\begin{table}[htbp]
\centering
\caption{Statistics of the Fine-tuning Datasets}
\label{tab:dataset}
\resizebox{\linewidth}{!}{
\begin{tabular}{lcc|cccc}
\toprule
\textbf{Dataset} & \textbf{\# Samples} & \textbf{\% Tool Calls} & \textbf{\# Total Msgs} & \textbf{\# Avg Msgs/Sample} & \textbf{\# Tool Msgs} & \textbf{\# Assist. Tool Msgs} \\
\midrule
\texttt{dataset\_tracker} & 1,132 & 428 (37.8\%) & 9,007 & 7.96 & 1,269 & 1,393 \\
\texttt{dataset\_mix} & 1,985 & 697 (35.1\%) & 15,064 & 7.59 & 2,166 & 2,340 \\
\bottomrule
\end{tabular}
}
\end{table}

\subsection{Baselines}
To evaluate the effectiveness of \toolname, we compare it against state-of-the-art approaches, categorized into:

\textbf{Category A: Vanilla LLMs.}
We compare against vanilla versions of leading general-purpose LLMs: \textit{Claude-Sonnet-4.5}, \textit{GPT-5-Mini}, and \textit{GPT-5.1}\footnote{The model versions used are \textit{claude-sonnet-4-5-20250929}, \textit{gpt-5-mini-2025-08-07}, and \textit{gpt-5.1-2025-11-13}.}. These models are prompted with standard instructions to generate Solidity code.

\textbf{Category B: Agent Frameworks.}
We compare against advanced agentic frameworks:
\begin{itemize}[leftmargin=*]
    \item GitHub Copilot~\cite{Github2025Github}: An AI-powered coding assistant that helps you write code faster by suggesting whole lines or functions directly inside your editor.
    \item DeepCode~\cite{Li2025DeepCode}: A code-specific agent framework focusing on repository-level generation.
    \item MetaGPT~\cite{Hong2024MetaGPT}: A multi-agent framework that assigns roles (e.g., Product Manager, Architect, Engineer) to LLMs.
    \item Qwen-Agent~\cite{QwenTeam2025QwenAgent}: An agent framework built upon the Qwen model family.
\end{itemize}

\textbf{Category C: Ablated Versions.}
To understand the contribution of each component in \toolname, we evaluate:
\begin{itemize}[leftmargin=*]
    \item \textbf{w/o Forge}: Removing the correctness feedback from Forge~\cite{FoundryContributors2025Foundry}.
    \item \textbf{w/o Slither}: Removing the security feedback from Slither~\cite{Feist2019Slither}.
    \item \textbf{w/o Tools}: Removing the file system tools, restricting the agent to memory-only context.
\end{itemize}

\textbf{Category D: Distilled Models.}
We compare our two distilled Qwen3-8B models, named solagent-tracker and solagent-mix (fine-tuned on \texttt{dataset\_tracker} and \texttt{dataset\_mix}, respectively), against the base Qwen3-8B model and a larger baseline (Qwen3-32B).

\subsection{Evaluation Metrics}
We use three key metrics to measure performance:

\textbf{Metric 1: Functional Correctness.}
We evaluate functional correctness using the Pass@$k$ metric, which has been widely used to assess the success rate of code generation models that meets specified requirements~\cite{Chen2021Evaluating,Yu2024CoderEval,Daspe2024Benchmarking}. This metric estimates the probability that at least one out of $k$ generated code samples passes the unit tests. Given $n$ samples generated for each problem, and $c$ samples that pass all unit tests, the unbiased estimator for Pass@$k$ is calculated as:
\begin{equation}
    Pass@k := \mathop{\mathbb{E}}\left[1 - \frac{\binom{n-c}{k}}{\binom{n}{k}}\right]
\end{equation}
In our experiments, we primarily report \textbf{Pass@1} (where $k=1$), which reflects the model's ability to generate correct code in a single attempt. Note that this overall Pass@1 is calculated across all test cases and all files. To further evaluate the quality of the generated logic independently of compilation issues, we also report the \textbf{Pass@1 (Mean $\pm$ Std)}. This metric is calculated by first computing the Pass@1 for each \textit{successfully compiled} contract file, and then reporting the mean and standard deviation across these compiled files. This provides a more nuanced view of the model's logic-building capability once the code is syntactically correct.
Additionally, inspired by previous research ~\cite{Peng2025SolEval,Inala2022Faultaware}, we report the \textit{Compilation Rate} to better fit the file-level smart contract generation task, defined as the percentage of generated files that successfully compile:
\begin{equation}
    Rate_{compile} = \frac{1}{N} \sum_{i=1}^{N} \mathbb{1}(\text{Compile}(s_i))
\end{equation}
where $N$ is the total number of generated smart contracts, $s_i$ is the $i$-th generated contract, and $\mathbb{1}(\text{Compile}(s_i))$ is an indicator function that equals 1 if $s_i$ compiles successfully, and 0 otherwise.

\textbf{Metric 2: Gas Efficiency.}
For correctly generated contracts, we calculate the average gas consumption of the passed test cases. In addition to comparing with baseline approaches, we also compare this against the gas usage of the original, human-written code in the repository. Lower gas usage indicates better optimization. 

\textbf{Metric 3: Security Vulnerabilities.}
We use Slither to detect security issues in the generated code. We report the total count of \textbf{High}, \textbf{Medium}, and \textbf{Low} severity vulnerabilities. In addition to comparing with baseline approaches, we also benchmark against the original, human-written code in the repository. A lower count indicates more secure code.

\subsection{Experimental Setting}\label{sec:exp-setting}
\noindent
\textbf{Implementation.} 
We implemented \toolname by building upon the MS-Agent framework~\cite{Li2023ModelScopeagent}, a general and customizable agent framework originally designed for real-world AI applications such as Python script generation and web automation. 
MS-Agent provides a user-friendly system library with seamless integration of model APIs and common APIs, along with native support for tool registration and memory control.
We selected MS-Agent over other agent frameworks due to its lightweight architecture, high customizability, and flexible callback system, which are essential for implementing our iterative refinement workflow.

However, MS-Agent lacks the specialized capabilities required for secure smart contract generation.
To adapt it to the rigorous requirements of Solidity development, we conducted extensive architectural modifications:
\textbf{(1) Domain-Specific Tool System.} We developed a custom file system tool with security sandbox constraints to prevent unauthorized access to original solutions and test files, while enabling contextual dependency exploration. The tool implements intelligent filtering to reduce context noise.
\textbf{(2) Multi-Dimensional Evaluation Loop.} We integrated Forge (compiler and testing) and Slither (static analysis) directly into the callback system to provide real-time feedback on functional correctness, security vulnerabilities, and gas efficiency.
\textbf{(3) Intelligent Termination Mechanism.} We designed a sophisticated stopping algorithm with over 10 termination conditions. Specifically, we set the maximum refinement rounds to 50. The process terminates early if the code passes all tests and security checks, if the pass rate stagnates for $N=2$ consecutive rounds, or if the feedback similarity between consecutive rounds exceeds $\tau=0.9$ (indicating a loop).
\textbf{(4) Best Code Tracking.} We implemented a state management system that dynamically tracks the optimal solution across iterations, ensuring quality even if later refinement rounds degrade performance.
\textbf{(5) Checkpoint \& Resume Support.} We developed a comprehensive checkpointing system that serializes the agent's state to disk, enabling seamless task resumption with idempotency guarantees.
\textbf{(6) Token Optimization Strategy.} We designed an intelligent message pruning strategy that reduces token consumption by approximately 60\% in long refinement sessions. 

\noindent
\textbf{Hardware \& Training.}
We utilized distinct hardware environments for training and inference to optimize resource usage.
For fine-tuning, we used a server equipped with 8 $\times$ Huawei Ascend 910B2 NPUs (64GB). The training process was implemented using the ms-swift library~\cite{Zhao2025SWIFT}, a scalable lightweight infrastructure for fine-tuning. We performed full-parameter fine-tuning on the Qwen3-8B model with a learning rate of $2e-5$, a batch size of 1, and gradient accumulation steps of 2. The model was trained for 3 epochs using the AdamW optimizer.
For inference and evaluation, we deployed the models on a cluster of 10 $\times$ NVIDIA V100 GPUs (32GB).

\noindent
\textbf{Summary Generation.}
For the summary-augmented portion of \texttt{dataset\_mix}, we used GPT-5-Mini to summarize the detailed method comments in the SolEval+ into concise functional requirements. 
These summarized requirements were then used as user queries to prompt the baseline models to generate new interaction trajectories, which were mixed with the original data.

\section{Results}
\label{sec:results}

To investigate the effectiveness of \toolname in generating correct, secure and gas-efficient smart contracts, we formulate the following three research questions:

\begin{itemize}[leftmargin=*]
    \item \textbf{RQ-1 Effectiveness Comparison.} \textit{How does \toolname perform compared to state-of-the-art LLMs and agent frameworks in terms of functional correctness, gas efficiency, and security?}
    \item \textbf{RQ-2 Ablation Study.} \textit{What is the contribution of each component (Forge feedback, Slither feedback, and file system tools) to the overall performance of \toolname?}
    \item \textbf{RQ-3 Distillation Effectiveness.} \textit{Can the capabilities of the multi-agent system be effectively distilled into a smaller model (Qwen3-8B) using the generated trajectories?}
\end{itemize}

\subsection{RQ-1 Effectiveness Comparison}
\label{sec:rq1}

\noindent
\textbf{Objective.}
We aim to evaluate whether \toolname outperforms general-purpose LLMs and existing agents in generating high-quality smart contracts. Specifically, we assess performance across three dimensions: functional correctness (Pass@1), gas efficiency, and security vulnerability reduction.

\noindent
\textbf{Experimental Design.}
We compare \toolname against two categories of baselines:
(1) \textbf{General-purpose LLMs}: Claude-Sonnet-4.5, GPT-5-Mini, and GPT-5.1, prompted directly with the requirements.
(2) \textbf{Agent Frameworks}: Copilot, DeepCode, MetaGPT, and Qwen-Agent, configured with the same base models.
We use the SolEval+ Benchmark (81 files, 1,188 test cases) for evaluation.
For functional correctness, we report the \textbf{Compilation Rate} and \textbf{Pass@1} (percentage of passed test cases).
For gas efficiency, we calculate the pairwise gas usage ratio between \toolname and baselines on common passed tests.
For security, we compare the number of vulnerabilities detected by Slither against the original human-written code (Baseline Repo).

\noindent
\textbf{Results.}
The results for functional correctness are presented in Table~\ref{tab:rq1-pass}.
We observe that:
(1) \textbf{\toolname significantly outperforms all baselines.}
With Claude-Sonnet-4.5, \toolname achieves a Pass@1 of \textbf{64.39\%}, surpassing the vanilla LLM (25.59\%) and the best agent baseline (Qwen-Agent, 28.37\%) by a large margin.
Similar trends are observed with GPT-5-Mini and GPT-5.1, where \toolname achieves 56.65\% and 54.71\% Pass@1, respectively.
(2) \textbf{High Compilation Rate.}
\toolname consistently achieves over 90\% compilation rate (e.g., 95.06\% with Claude-Sonnet-4.5), whereas vanilla baselines often struggle to generate compilable code (typically 30\%-45\%).
(3) \textbf{Stability.}
The mean Pass@1 per file (e.g., 0.7795 $\pm$ 0.2941 for Claude-Sonnet-4.5) demonstrates both high performance and low variance, indicating stable performance across different contracts.

\begin{table}[htbp]
  \centering
  \caption{Functional Correctness (Pass@1) and Compilation Rate Comparison}
  \label{tab:rq1-pass}
  \resizebox{.85\linewidth}{!}{
    \begin{tabular}{llccccc}
    \toprule
    \textbf{Source} & \textbf{Model} & \textbf{Files} & \makecell{\textbf{Compile}\\\textbf{Rate}} & \makecell{\textbf{Passed}\\\textbf{Tests}} & \textbf{Pass@1} & \makecell{\textbf{Pass@1}\\\textbf{(Mean$\pm$Std)}} \\
    \midrule
    Baseline (Repo) & N/A & 81/81 & 100.00\% & 1188 & 100.00\% & 1.0000 $\pm$ 0.0000 \\
    \midrule
    \multirow{3}{*}{Vanilla LLM} & Claude-Sonnet-4.5 & 32/81 & 39.51\% & 304 & 25.59\% & 0.7743 $\pm$ 0.3142 \\
    & GPT-5-Mini & 26/81 & \underline{32.10\%} & 167 & \underline{14.06\%} & 0.6817 $\pm$ 0.3394 \\
    & GPT-5.1 & 33/81 & 40.74\% & 217 & 18.27\% & 0.6586 $\pm$ 0.3074 \\
    \specialrule{1pt}{1pt}{1pt}
    \multirow{3}{*}{Copilot} & Claude-Sonnet-4.5 & 26/81 & 32.10\% & 119 & 10.02\% & 0.7845 $\pm$ 0.3404 \\
    & GPT-5-Mini & 31/81 & 38.27\% & 250 & 21.04\% & 0.7043 $\pm$ 0.3253 \\
    & GPT-5.1 & 27/81 & 33.33\% & 120 & 10.10\% & 0.6673 $\pm$ 0.3390 \\
    \cmidrule(lr){2-7}
    \multirow{3}{*}{DeepCode} & Claude-Sonnet-4.5 & 30/81 & 37.04\% & 161 & 13.55\% & 0.7449 $\pm$ 0.3006 \\
    & GPT-5-Mini & 29/81 & 35.80\% & 206 & 17.34\% & 0.6762 $\pm$ 0.3158 \\
    & GPT-5.1 & 33/81 & 40.74\% & 235 & 19.78\% & 0.5583 $\pm$ 0.3554 \\
    \cmidrule(lr){2-7}
    \multirow{3}{*}{MetaGPT} & Claude-Sonnet-4.5 & 29/81 & 35.80\% & 140 & 11.78\% & 0.7368 $\pm$ 0.3105 \\
    & GPT-5-Mini & 23/81 & \underline{28.40\%} & 115 & \underline{9.68\%} & 0.6396 $\pm$ 0.3449 \\
    & GPT-5.1 & 28/81 & 34.57\% & 131 & 11.03\% & 0.5938 $\pm$ 0.3420 \\
    \cmidrule(lr){2-7}
    \multirow{3}{*}{Qwen-Agent} & Claude-Sonnet-4.5 & 37/81 & 45.68\% & 337 & 28.37\% & 0.7278 $\pm$ 0.3242 \\
    & GPT-5-Mini & 30/81 & 37.04\% & 239 & 20.12\% & 0.7519 $\pm$ 0.2959 \\
    & GPT-5.1 & 30/81 & 37.04\% & 144 & 12.12\% & 0.6823 $\pm$ 0.3392 \\
    
    \midrule
    \multirow{3}{*}{\makecell[l]{\toolnameb\\\textbf{Summary}}} & Claude-Sonnet-4.5 & 71/81 & \textbf{87.65\%} & 566 & 47.64\% & 0.7791 $\pm$ 0.3166 \\
    & GPT-5-Mini & 62/81 & 76.54\% & 614 & \textbf{51.68\%} & 0.7819 $\pm$ 0.2984 \\
    & GPT-5.1 & 66/81 & 81.48\% & 394 & 33.16\% & 0.6342 $\pm$ 0.3271 \\
    \midrule
    \multirow{3}{*}{\toolnameb} & \cellcolor{lightgray}Claude-Sonnet-4.5 & \cellcolor{lightgray}77/81 & \cellcolor{lightgray}\textbf{95.06\%} & \cellcolor{lightgray}765 & \cellcolor{lightgray}\textbf{64.39\%} & \cellcolor{lightgray}0.7795 $\pm$ 0.2941 \\
    & \cellcolor{lightgray}GPT-5-Mini & \cellcolor{lightgray}73/81 & \cellcolor{lightgray}90.12\% & \cellcolor{lightgray}673 & \cellcolor{lightgray}56.65\% & \cellcolor{lightgray}0.7642 $\pm$ 0.2873 \\
    & \cellcolor{lightgray}GPT-5.1 & \cellcolor{lightgray}74/81 & \cellcolor{lightgray}91.36\% & \cellcolor{lightgray}650 & \cellcolor{lightgray}54.71\% & \cellcolor{lightgray}0.6808 $\pm$ 0.3010 \\
    \bottomrule
    \end{tabular}%
  }
\end{table}

\noindent
\underline{\textbf{Gas Efficiency Analysis.}}
Table~\ref{tab:rq1-gas} presents the comprehensive gas efficiency analysis, including both pairwise test case comparison and file-level aggregation.
We observe that \toolname's gas efficiency varies across different baselines and models. Results show that \toolname achieves competitive gas efficiency against Claude-based models (mean ratios near 1.0), but consumes more gas when compared to GPT-5-Mini variants (mean ratios 1.44-2.39). This trade-off is expected, as our iterative refinement process prioritizes functional correctness and security over gas optimization.
However, the trimmed mean ratios (Trim5\%) remain close to 1.0, indicating that the majority of test cases exhibit reasonable gas efficiency.
At the file level, results show that \toolname achieves better gas efficiency in a substantial proportion of files across different baselines.

\begin{table}[htbp]
  \centering
  \caption{Gas Efficiency Analysis: Pairwise Test Case and File-Level Comparison}
  \label{tab:rq1-gas}
  \resizebox{\linewidth}{!}{
    \begin{tabular}{llccccccc}
    \toprule
    & & \multicolumn{4}{c}{\textbf{Pairwise Test Case Comparison}} & \multicolumn{3}{c}{\textbf{File-Level Comparison}} \\
    \cmidrule(lr){3-6} \cmidrule(lr){7-9}
    \textbf{\toolnameb vs.} & \textbf{Model} & \textbf{\# Common} & \textbf{Mean Ratio} & \textbf{Trim5\%} & \textbf{P90} & \textbf{\# Files} & \textbf{Better} & \textbf{Worse} \\
    \midrule
    \multirow{3}{*}{Vanilla LLM} & Claude-Sonnet-4.5 & 288 & \textbf{0.9958} & \textbf{0.9985} & 1.0005 & 30 & \textbf{10} & 8 \\
    & GPT-5-Mini & 161 & 2.3937 & 1.3778 & 3.7088 & 24 & 6 & 14 \\
    & GPT-5.1 & 211 & 1.6282 & \textbf{0.9724} & 1.1168 & 32 & 10 & 11 \\
    \specialrule{1pt}{1pt}{1pt}
    \multirow{3}{*}{Copilot} & Claude-Sonnet-4.5 & 115 & 1.0201 & \textbf{0.9998} & 1.0017 & 25 & \textbf{10} & 4 \\
    & GPT-5-Mini & 243 & 1.4462 & 1.0148 & 1.1785 & 29 & 9 & 16 \\
    & GPT-5.1 & 105 & 1.0296 & 1.0058 & 1.0771 & 24 & \textbf{11} & 6 \\
    \cmidrule(lr){2-9}
    \multirow{3}{*}{DeepCode} & Claude-Sonnet-4.5 & 148 & 1.0021 & \textbf{0.9958} & 1.0106 & 29 & \textbf{10} & 8 \\
    & GPT-5-Mini & 201 & 1.4534 & \textbf{0.9969} & 1.0789 & 28 & \textbf{13} & 10 \\
    & GPT-5.1 & 150 & 1.0979 & \textbf{0.9958} & 1.0971 & 29 & \textbf{13} & 10 \\
    \cmidrule(lr){2-9}
    \multirow{3}{*}{MetaGPT} & Claude-Sonnet-4.5 & 135 & \textbf{0.9950} & \textbf{0.9987} & 1.0075 & 29 & 9 & 11 \\
    & GPT-5-Mini & 104 & 1.0671 & 1.0217 & 1.1746 & 21 & 8 & 11 \\
    & GPT-5.1 & 101 & 1.0197 & 1.0020 & 1.0302 & 24 & 5 & 9 \\
    \cmidrule(lr){2-9}
    \multirow{3}{*}{Qwen-Agent} & Claude-Sonnet-4.5 & 314 & 1.0052 & \textbf{0.9985} & 1.0023 & 35 & \textbf{14} & 9 \\
    & GPT-5-Mini & 229 & 1.4384 & 1.0036 & 1.1359 & 29 & \textbf{13} & 10 \\
    & GPT-5.1 & 122 & 1.0188 & 1.0024 & 1.0737 & 27 & \textbf{11} & 8 \\
    \bottomrule
    \end{tabular}%
  }
  \vspace{0.5ex}
  \footnotesize
  \raggedright
  \textit{Note:} \textbf{Pairwise Comparison}: Ratios < 1.0 indicate \toolnameb is more efficient. \textbf{\# Common}: Common passed test cases. \textbf{Trim5\%}: 5\% trimmed mean. \textbf{P90}: 90th percentile.
  \textbf{File-Level Comparison}: \textbf{\# Files}: Files with common passed test cases. \textbf{Better/Worse}: Number of files where \toolnameb consumes less/more total gas.
\end{table}

\noindent
\underline{\textbf{Security Analysis.}}
Table~\ref{tab:rq1-vuln} presents the vulnerability comparison.
\toolname demonstrates a strong capability in generating secure code.
Compared to the original human-written code (Baseline Repo), \toolname (Claude) reduces the total number of vulnerabilities by \textbf{15.70\%} (247 vs 293). This improvement is even more pronounced with GPT-5-Mini, achieving a \textbf{39.77\%} reduction.
This indicates that the integration of Slither feedback effectively guides the agent to fix security issues during the refinement process.
It is worth noting that some baselines also exhibit significant reductions in vulnerabilities (e.g., Vanilla LLM GPT-5-Mini with -57.14\%).
However, this is primarily attributed to their lower compilation rates and fewer passed test cases, which results in a smaller comparison base.
Additionally, these models often generate code that, while potentially having fewer detected vulnerabilities, fails to meet functional requirements (i.e., fails tests).

\begin{table}[htbp]
  \centering
  \caption{Vulnerability Comparison against Baseline (Repo)}
  \label{tab:rq1-vuln}
  \resizebox{.85\linewidth}{!}{
    \begin{tabular}{llccccc}
    \toprule
    \textbf{Method} & \textbf{Model} & \makecell{\textbf{\# Common}\\\textbf{Files}} & \makecell{\textbf{Baseline}\\\textbf{Vuln}} & \makecell{\textbf{Method}\\\textbf{Vuln}} & \textbf{$\Delta$\% vs Base} & \makecell{\textbf{Vuln}\\\textbf{Diff}} \\
    \midrule
    \multirow{3}{*}{Vanilla LLM} & Claude-Sonnet-4.5 & 32 & 66 & 56 & -15.15\% & -10 \\
    & GPT-5-Mini & 26 & 49 & 21 & -57.14\% & -28 \\
    & GPT-5.1 & 33 & 63 & 36 & -42.86\% & -27 \\
    \specialrule{1pt}{1pt}{1pt}
    \multirow{3}{*}{Copilot} & Claude-Sonnet-4.5 & 26 & 26 & 30 & +15.38\% & +4 \\
    & GPT-5-Mini & 31 & 72 & 40 & -44.44\% & -32 \\
    & GPT-5.1 & 27 & 63 & 27 & -57.14\% & -36 \\
    \cmidrule(lr){2-7}
    \multirow{3}{*}{DeepCode} & Claude-Sonnet-4.5 & 30 & 56 & 41 & -26.79\% & -15 \\
    & GPT-5-Mini & 29 & 69 & 41 & -40.58\% & -28 \\
    & GPT-5.1 & 33 & 57 & 32 & -43.86\% & -25 \\
    \cmidrule(lr){2-7}
    \multirow{3}{*}{MetaGPT} & Claude-Sonnet-4.5 & 29 & 40 & 30 & -25.00\% & -10 \\
    & GPT-5-Mini & 23 & 35 & 20 & -42.86\% & -15 \\
    & GPT-5.1 & 28 & 55 & 34 & -38.18\% & -21 \\
    \cmidrule(lr){2-7}
    \multirow{3}{*}{Qwen-Agent} & Claude-Sonnet-4.5 & 37 & 47 & 55 & +17.02\% & +8 \\
    & GPT-5-Mini & 30 & 58 & 22 & -62.07\% & -36 \\
    & GPT-5.1 & 30 & 57 & 27 & -52.63\% & -30 \\
    \midrule
    \multirow{3}{*}{\makecell[l]{\toolnameb\\\textbf{Summary}}} & Claude-Sonnet-4.5 & 71 & 232 & 293 & +26.29\% & +61 \\
    & GPT-5-Mini & 62 & 188 & 118 & -37.23\% & -70 \\
    & GPT-5.1 & 66 & 182 & 228 & +25.27\% & +46 \\
    \midrule
    \multirow{3}{*}{\toolnameb} & \cellcolor{lightgray}Claude-Sonnet-4.5 & \cellcolor{lightgray}77 & \cellcolor{lightgray}293 & \cellcolor{lightgray}247 & \cellcolor{lightgray}-15.70\% & \cellcolor{lightgray}-46 \\
    & \cellcolor{lightgray}GPT-5-Mini & \cellcolor{lightgray}73 & \cellcolor{lightgray}259 & \cellcolor{lightgray}156 & \cellcolor{lightgray}-39.77\% & \cellcolor{lightgray}-103 \\
    & \cellcolor{lightgray}GPT-5.1 & \cellcolor{lightgray}74 & \cellcolor{lightgray}242 & \cellcolor{lightgray}182 & \cellcolor{lightgray}-24.79\% & \cellcolor{lightgray}-60 \\
    \bottomrule
    \end{tabular}%
  }
\end{table}

\begin{figure*}[htbp]
  \centering
  \includegraphics[width=.95\linewidth]{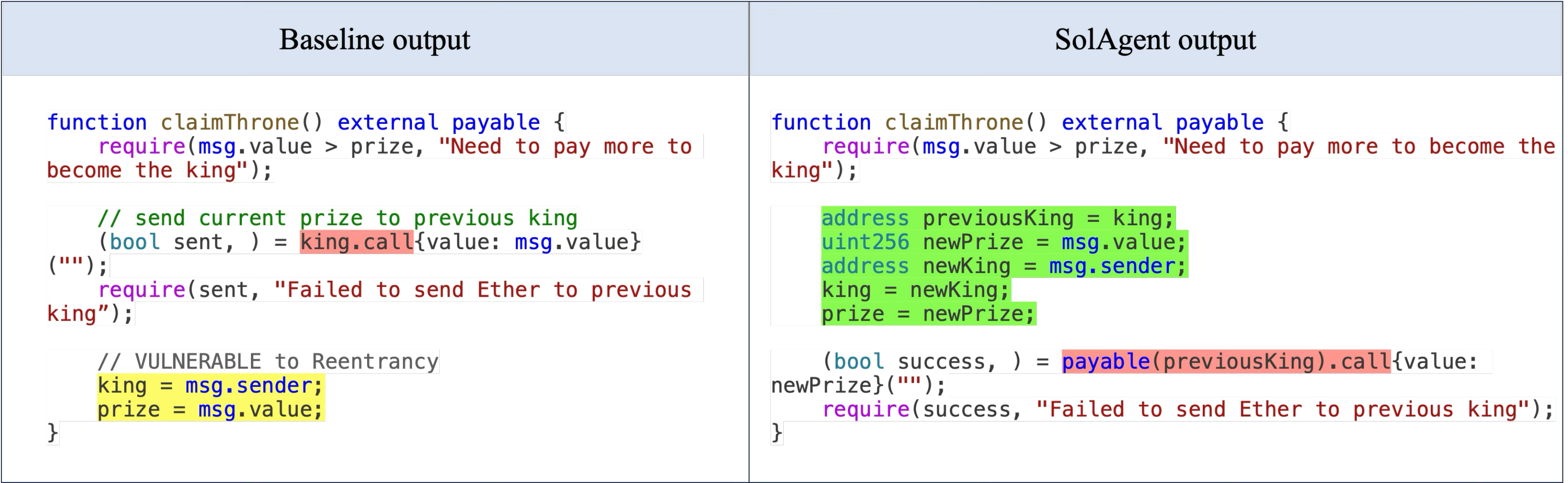}
  \caption{Example: Reentrancy fix in \texttt{claimThrone} by \toolname versus a baseline.}
  \label{fig:eg-vuln}
\end{figure*}

As an example of vulnerability remediation, Figure~\ref{fig:eg-vuln} compares the \texttt{claimThrone} logic produced by a baseline and by \toolname.
In the baseline code, \texttt{king} is updated only after the external call \verb|king.call{value: msg.value}("")|.
This ordering is vulnerable to reentrancy: if the current \texttt{king} is a malicious contract that re-enters \texttt{claimThrone} in its fallback, it can drain the contract balance by repeatedly claiming the throne before \texttt{king} and \texttt{prize} are updated.
\toolname, guided by Slither feedback, instead follows the checks-effects-interactions pattern: it first caches \texttt{previousKing = king} and \texttt{newKing = msg.sender}, updates \texttt{king} and \texttt{prize} to the new values, and only then performs the external transfer to \texttt{previousKing}.
Thus, state is consistent before any external call, eliminating the reentrancy window.

\noindent
\underline{\textbf{Code Complexity Analysis.}}
Table~\ref{tab:rq1-complexity} presents the comprehensive statistics on Lines of Code (LOC) and cyclomatic complexity.
Notably, \toolname generates code with higher LOC and complexity compared to baselines.
For instance, with Claude-Sonnet-4.5, \toolname produces an average of 216.4 LOC per file with an average complexity of 22.2, compared to the vanilla LLM's 148.7 LOC and 14.2 complexity.
This increase directly correlates with \toolname's superior Pass@1 rate (64.39\% vs 25.59\%).
The higher complexity reflects more complete requirement implementation: \toolname generates comprehensive solutions with proper error handling, edge case coverage, and security checks, which are necessary to pass the rigorous test suites in SolEval+.
In contrast, baselines often generate simpler but incomplete code that fails to satisfy functional requirements, resulting in lower Pass@1 despite having lower complexity metrics.

\begin{table}[htbp]
  \centering
  \caption{Code Complexity: LOC and Cyclomatic Complexity Comparison}
  \label{tab:rq1-complexity}
  \resizebox{\linewidth}{!}{
    \begin{tabular}{llccccccc}
    \toprule
    \textbf{Method} & \textbf{Model} & \textbf{Files} & \textbf{LOC} & \textbf{Avg LOC} & \textbf{PLOC} & \textbf{Avg PLOC} & \textbf{Complexity} & \textbf{Avg Complexity} \\
    \midrule
    \multirow{3}{*}{Vanilla LLM} & Claude-Sonnet-4.5 & 32 & 4759 & 148.7 & 10471 & 327.2 & 455 & 14.2 \\
    & GPT-5-Mini & 26 & 3411 & 131.2 & 6179 & 237.7 & 509 & 19.6 \\
    & GPT-5.1 & 33 & 7802 & 236.4 & 12031 & 364.6 & 624 & 18.9 \\
    \specialrule{1pt}{1pt}{1pt}
    \multirow{3}{*}{Copilot} & Claude-Sonnet-4.5 & 26 & 3061 & 117.7 & 5946 & 228.7 & 372 & 14.3 \\
    & GPT-5-Mini & 31 & 4599 & 148.4 & 7371 & 237.8 & 721 & 23.3 \\
    & GPT-5.1 & 27 & 4549 & 168.5 & 8740 & 323.7 & 414 & 15.3 \\
    \cmidrule(lr){2-9}
    \multirow{3}{*}{DeepCode} & Claude-Sonnet-4.5 & 30 & 4196 & 139.9 & 11680 & 389.3 & 511 & 17.0 \\
    & GPT-5-Mini & 29 & 3647 & 125.8 & 5960 & 205.5 & 673 & 23.2 \\
    & GPT-5.1 & 33 & 6568 & 199.0 & 9626 & 291.7 & 533 & 16.2 \\
    \cmidrule(lr){2-9}
    \multirow{3}{*}{MetaGPT} & Claude-Sonnet-4.5 & 29 & 4365 & 150.5 & 7137 & 246.1 & 436 & 15.0 \\
    & GPT-5-Mini & 23 & 2406 & 104.6 & 3747 & 162.9 & 284 & 12.3 \\
    & GPT-5.1 & 28 & 5392 & 192.6 & 9609 & 343.2 & 543 & 19.4 \\
    \cmidrule(lr){2-9}
    \multirow{3}{*}{Qwen-Agent} & Claude-Sonnet-4.5 & 37 & 6580 & 177.8 & 14718 & 397.8 & 648 & 17.5 \\
    & GPT-5-Mini & 30 & 4203 & 140.1 & 8253 & 275.1 & 654 & 21.8 \\
    & GPT-5.1 & 30 & 6602 & 220.1 & 10197 & 339.9 & 500 & 16.7 \\
    \midrule
    \multirow{3}{*}{\toolnameb} & \cellcolor{lightgray}Claude-Sonnet-4.5 & \cellcolor{lightgray}77 & \cellcolor{lightgray}16661 & \cellcolor{lightgray}216.4 & \cellcolor{lightgray}28154 & \cellcolor{lightgray}365.6 & \cellcolor{lightgray}1708 & \cellcolor{lightgray}22.2 \\
    & \cellcolor{lightgray}GPT-5-Mini & \cellcolor{lightgray}73 & \cellcolor{lightgray}12663 & \cellcolor{lightgray}173.5 & \cellcolor{lightgray}20864 & \cellcolor{lightgray}285.8 & \cellcolor{lightgray}2071 & \cellcolor{lightgray}28.4 \\
    & \cellcolor{lightgray}GPT-5.1 & \cellcolor{lightgray}74 & \cellcolor{lightgray}18910 & \cellcolor{lightgray}255.5 & \cellcolor{lightgray}25557 & \cellcolor{lightgray}345.4 & \cellcolor{lightgray}1803 & \cellcolor{lightgray}24.4 \\
    \bottomrule
    \end{tabular}%
  }
  \vspace{0.5ex}
  \footnotesize
  \raggedright
  \textit{Note:} \textbf{LOC} excludes comments/blanks; \textbf{PLOC} denotes physical LOC, which includes all lines. \textbf{Cyclomatic Complexity} = Number of decision nodes + 1 (including: if, while, for, case, catch, \&\&, ||, ?, require, assert).
\end{table}

\noindent
\underline{\textbf{Token Analysis.}}
Tables~\ref{tab:rq1-token} and~\ref{tab:rq1-token-avg} present the comprehensive token consumption statistics on compiled files for all methods, broken down by Coding and Refine phases (total counts and per-file/per-round averages).
While \toolname consumes more tokens than most baselines due to its iterative refinement approach, this overhead translates directly into superior code quality.
With Claude-Sonnet-4.5, \toolname consumes 5,363 prompt tokens and 5,545 completion tokens per file during the initial coding phase, comparable to the vanilla LLM's 4,807 and 4,211.
The refinement phase adds significant overhead: 4.6M prompt tokens and 887K completion tokens for 77 files, resulting in averages of 65,510 prompt tokens and 17,064 completion tokens per file.
Compared to MetaGPT (137,560 prompt tokens per file for single-pass generation), \toolname uses significantly fewer total tokens despite multiple refinement rounds. While each refinement round incurs higher prompt costs due to tool call interactions (23,794 tokens per round), the cumulative overhead remains lower than MetaGPT's upfront cost.
This investment yields a 5.5$\times$ improvement in Pass@1 (64.39\% vs 11.78\%), demonstrating favorable cost-effectiveness.

\begin{table}[htbp]
  \centering
  \caption{Token Consumption Statistics: Total Token Counts}
  \label{tab:rq1-token}
  \resizebox{\linewidth}{!}{
    \begin{tabular}{llcccccccc}
    \toprule
    \textbf{Method} & \textbf{Model} & \makecell{\textbf{Coding}\\\textbf{Prompt}} & \makecell{\textbf{Coding}\\\textbf{Completion}} & \makecell{\textbf{Refine}\\\textbf{Prompt}} & \makecell{\textbf{Refine}\\\textbf{Completion}} & \makecell{\textbf{Total}\\\textbf{Prompt}} & \makecell{\textbf{Total}\\\textbf{Completion}} \\
    \midrule
    \multirow{3}{*}{Vanilla LLM} & Claude-Sonnet-4.5 & 153,835 & 134,748 & - & - & 153,835 & 134,748 \\
    & GPT-5-Mini & 83,342 & 92,827 & - & - & 83,342 & 92,827 \\
    & GPT-5.1 & 161,625 & 103,164 & - & - & 161,625 & 103,164 \\
    \specialrule{1pt}{1pt}{1pt}
    \multirow{3}{*}{MetaGPT} & Claude-Sonnet-4.5 & 3,989,233 & 241,933 & - & - & 3,989,233 & 241,933 \\
    & GPT-5-Mini & 3,381,285 & 403,427 & - & - & 3,381,285 & 403,427 \\
    & GPT-5.1 & 679,060 & 584,938 & - & - & 679,060 & 584,938 \\
    \cmidrule(lr){2-8}
    \multirow{3}{*}{DeepCode} & Claude-Sonnet-4.5 & 164,159 & 155,059 & - & - & 164,159 & 155,059 \\
    & GPT-5-Mini & 105,171 & 394,997 & - & - & 105,171 & 394,997 \\
    & GPT-5.1 & 182,271 & 96,171 & - & - & 182,271 & 96,171 \\
    \cmidrule(lr){2-8}
    \multirow{3}{*}{Qwen-Agent} & Claude-Sonnet-4.5 & 180,188 & 187,571 & - & - & 180,188 & 187,571 \\
    & GPT-5-Mini & 101,230 & 131,050 & - & - & 101,230 & 131,050 \\
    & GPT-5.1 & 134,582 & 91,073 & - & - & 134,582 & 91,073 \\
    \midrule
    \multirow{3}{*}{\toolnameb} & \cellcolor{lightgray}Claude-Sonnet-4.5 & \cellcolor{lightgray}412,959 & \cellcolor{lightgray}426,960 & \cellcolor{lightgray}4,631,282 & \cellcolor{lightgray}886,931 & \cellcolor{lightgray}5,044,241 & \cellcolor{lightgray}1,313,891 \\
    & \cellcolor{lightgray}GPT-5-Mini & \cellcolor{lightgray}322,934 & \cellcolor{lightgray}393,409 & \cellcolor{lightgray}2,254,720 & \cellcolor{lightgray}652,701 & \cellcolor{lightgray}2,577,654 & \cellcolor{lightgray}1,046,110 \\
    & \cellcolor{lightgray}GPT-5.1 & \cellcolor{lightgray}350,432 & \cellcolor{lightgray}263,847 & \cellcolor{lightgray}4,856,987 & \cellcolor{lightgray}848,589 & \cellcolor{lightgray}5,207,419 & \cellcolor{lightgray}1,112,436 \\
    \bottomrule
    \end{tabular}%
  }
\end{table}

\begin{table}[htbp]
  \centering
  \caption{Token Consumption Statistics: Per-File and Per-Round Averages}
  \label{tab:rq1-token-avg}
  \resizebox{\linewidth}{!}{
    \begin{tabular}{llcccccccc}
    \toprule
    \textbf{Method} & \textbf{Model} & \makecell{\textbf{Avg Coding}\\\textbf{Prompt/F}} & \makecell{\textbf{Avg Coding}\\\textbf{Completion/F}} & \makecell{\textbf{Avg Total}\\\textbf{Prompt/F}} & \makecell{\textbf{Avg Total}\\\textbf{Completion/F}} & \makecell{\textbf{Avg}\\\textbf{Prompt/R}} & \makecell{\textbf{Avg}\\\textbf{Completion/R}} \\
    \midrule
    \multirow{3}{*}{Vanilla LLM} & Claude-Sonnet-4.5 & 4,807 & 4,211 & 4,807 & 4,211 & - & - \\
    & GPT-5-Mini & 3,205 & 3,570 & 3,205 & 3,570 & - & - \\
    & GPT-5.1 & 4,898 & 3,126 & 4,898 & 3,126 & - & - \\
    \specialrule{1pt}{1pt}{1pt}
    \multirow{3}{*}{MetaGPT} & Claude-Sonnet-4.5 & 137,560 & 8,343 & 137,560 & 8,343 & - & - \\
    & GPT-5-Mini & 147,012 & 17,540 & 147,012 & 17,540 & - & - \\
    & GPT-5.1 & 24,252 & 20,891 & 24,252 & 20,891 & - & - \\
    \cmidrule(lr){2-8}
    \multirow{3}{*}{DeepCode} & Claude-Sonnet-4.5 & 5,472 & 5,169 & 5,472 & 5,169 & - & - \\
    & GPT-5-Mini & 3,627 & 13,621 & 3,627 & 13,621 & - & - \\
    & GPT-5.1 & 5,523 & 2,914 & 5,523 & 2,914 & - & - \\
    \cmidrule(lr){2-8}
    \multirow{3}{*}{Qwen-Agent} & Claude-Sonnet-4.5 & 4,870 & 5,069 & 4,870 & 5,069 & - & - \\
    & GPT-5-Mini & 3,374 & 4,368 & 3,374 & 4,368 & - & - \\
    & GPT-5.1 & 4,486 & 3,036 & 4,486 & 3,036 & - & - \\
    \midrule
    \multirow{3}{*}{\toolnameb} & \cellcolor{lightgray}Claude-Sonnet-4.5 & \cellcolor{lightgray}5,363 & \cellcolor{lightgray}5,545 & \cellcolor{lightgray}65,510 & \cellcolor{lightgray}17,064 & \cellcolor{lightgray}23,794 & \cellcolor{lightgray}6,198 \\
    & \cellcolor{lightgray}GPT-5-Mini & \cellcolor{lightgray}4,424 & \cellcolor{lightgray}5,389 & \cellcolor{lightgray}35,310 & \cellcolor{lightgray}14,330 & \cellcolor{lightgray}11,989 & \cellcolor{lightgray}4,866 \\
    & \cellcolor{lightgray}GPT-5.1 & \cellcolor{lightgray}4,736 & \cellcolor{lightgray}3,566 & \cellcolor{lightgray}70,371 & \cellcolor{lightgray}15,033 & \cellcolor{lightgray}16,637 & \cellcolor{lightgray}3,554 \\
    \bottomrule
    \end{tabular}%
  }
  \vspace{0.5ex}
  \footnotesize
  \raggedright
  \textit{Note:} F = File; R = Round (i.e., per refinement round). 
\end{table}

\intuition{
\textbf{Answer to RQ-1:}
\toolname significantly outperforms state-of-the-art LLMs and agents in Pass@1 (64.39\%).
It also generates more secure code, reducing vulnerabilities by up to 39.77\% compared to human-written baselines, while maintaining comparable gas efficiency.
The dual-loop refinement prioritizes correctness and security, effectively addressing the trade-off inherent in single-pass generation.
Pass@1, vulnerability counts, and gas efficiency together provide a multi-dimensional view of code quality.
}
\subsection{RQ-2 Ablation Study}
\label{sec:rq2}

\noindent
\textbf{Objective.}
To understand the contribution of each component in \toolname, we conduct an ablation study by removing:
(1) \textbf{Forge Feedback}: To assess its impact on functional correctness.
(2) \textbf{Slither Feedback}: To assess its impact on security.
(3) \textbf{File System Tools}: To assess the importance of context awareness.

\noindent
\textbf{Experimental Design.}
We compare \toolname against three ablated versions:
\begin{itemize}[leftmargin=*]
    \item \textbf{w/o Forge}: The agent receives no compilation or test feedback.
    \item \textbf{w/o Slither}: The agent receives no security vulnerability feedback.
    \item \textbf{w/o Tools}: The agent cannot access the file system to read dependencies or project structure.
\end{itemize}
All experiments are conducted using Claude-Sonnet-4.5, GPT-5-Mini, and GPT-5.1 as base models.

\begin{table}[htbp]
  \centering
  \caption{Ablation Study: Functional Correctness (Pass@1)}
  \label{tab:rq2-ablation}
  \resizebox{.85\linewidth}{!}{
    \begin{tabular}{llccccc}
    \toprule
    \textbf{Variant} & \textbf{Model} & \textbf{Files} & \makecell{\textbf{Compile}\\\textbf{Rate}} & \makecell{\textbf{Passed}\\\textbf{Tests}} & \textbf{Pass@1 (\%)} & \makecell{\textbf{Pass@1}\\\textbf{(Mean$\pm$Std)}} \\
    \midrule
    \rowcolor{lightgray} \toolname & Claude-Sonnet-4.5 & 77/81 & 95.06\% & 765 & 64.39\% & 0.7795 $\pm$ 0.2941 \\
    w/o Forge & Claude-Sonnet-4.5 & 32/81 & 39.51\% & 311 & 26.18\% & 0.7769 $\pm$ 0.3009 \\
    w/o Slither & Claude-Sonnet-4.5 & 73/81 & 90.12\% & 764 & 64.31\% & 0.7908 $\pm$ 0.2860 \\
    w/o Tools & Claude-Sonnet-4.5 & 71/81 & 87.65\% & 685 & 57.66\% & 0.7532 $\pm$ 0.3027 \\
    \midrule
    \rowcolor{lightgray} \toolname & GPT-5-Mini & 73/81 & 90.12\% & 673 & 56.65\% & 0.7642 $\pm$ 0.2873 \\
    w/o Forge & GPT-5-Mini & 34/81 & 41.98\% & 204 & 17.17\% & 0.6504 $\pm$ 0.3174 \\
    w/o Slither & GPT-5-Mini & 67/81 & 82.72\% & 641 & 53.96\% & 0.8192 $\pm$ 0.2543 \\
    w/o Tools & GPT-5-Mini & 69/81 & 85.19\% & 671 & 56.48\% & 0.7293 $\pm$ 0.3123 \\
    \midrule
    \rowcolor{lightgray} \toolname & GPT-5.1 & 74/81 & 91.36\% & 650 & 54.71\% & 0.6808 $\pm$ 0.3010 \\
    w/o Forge & GPT-5.1 & 38/81 & 46.91\% & 224 & 18.86\% & 0.6338 $\pm$ 0.3287 \\
    w/o Slither & GPT-5.1 & 67/81 & 82.72\% & 565 & 47.56\% & 0.6583 $\pm$ 0.3169 \\
    w/o Tools & GPT-5.1 & 61/81 & 75.31\% & 377 & 31.73\% & 0.6752 $\pm$ 0.3070 \\
    \bottomrule
    \end{tabular}%
  }
\end{table}

\begin{table}[htbp]
  \centering
  \caption{Ablation Study: Vulnerability Comparison}
  \label{tab:rq2-vuln}
  \resizebox{.85\linewidth}{!}{
    \begin{tabular}{llccccc}
    \toprule
    \textbf{Comparison} & \textbf{Model} & \makecell{\textbf{\# Common}\\\textbf{Files}} & \makecell{\textbf{\toolnameb}\\\textbf{Vuln}} & \makecell{\textbf{Ablation}\\\textbf{Vuln}} & \textbf{$\Delta$\%} & \makecell{\textbf{Vuln}\\\textbf{Diff}} \\
    \midrule
    w/o Forge & Claude-Sonnet-4.5 & 32 & 37 & 37 & +0.00\% & +0 \\
    \textbf{w/o Slither} & \textbf{Claude-Sonnet-4.5} & \textbf{73} & \textbf{236} & \textbf{273} & \textbf{+15.68\%} & \textbf{+37} \\
    w/o Tools & Claude-Sonnet-4.5 & 71 & 231 & 257 & +11.26\% & +26 \\
    \midrule
    w/o Forge & GPT-5-Mini & 34 & 25 & 35 & +40.00\% & +10 \\
    w/o Slither & GPT-5-Mini & 67 & 125 & 117 & -6.40\% & -8 \\
    w/o Tools & GPT-5-Mini & 67 & 140 & 137 & -2.14\% & -3 \\
    \midrule
    w/o Forge & GPT-5.1 & 38 & 41 & 47 & +14.63\% & +6 \\
    \textbf{w/o Slither} & \textbf{GPT-5.1} & \textbf{65} & \textbf{122} & \textbf{127} & \textbf{+4.10\%} & \textbf{+5} \\
    w/o Tools & GPT-5.1 & 61 & 85 & 72 & -15.29\% & -13 \\
    \bottomrule
    \end{tabular}%
  }
\end{table}

\begin{table}[htbp]
  \centering
  \caption{Ablation Study: Metrics at Min-Vuln Round (Prioritizing Security)}
  \label{tab:rq2-min-vuln}
  \resizebox{.85\linewidth}{!}{
    \begin{tabular}{llcccccc}
    \toprule
    \textbf{Variant} & \textbf{Model} & \textbf{Files} & \makecell{\textbf{Pass@1 (\%)}} & \makecell{\textbf{\toolnameb}\\\textbf{Vuln}} & \makecell{\textbf{Ablation}\\\textbf{Vuln}} & \textbf{$\Delta$\%} & \makecell{\textbf{Vuln}\\\textbf{Diff}} \\
    \midrule
    w/o Forge & Claude-Sonnet-4.5 & 32/81 & 26.18\% & 28 & 28 & +0.00\% & +0 \\
    \textbf{w/o Slither} & \textbf{Claude-Sonnet-4.5} & \textbf{73/81} & \textbf{62.37\%} & \textbf{202} & \textbf{250} & \textbf{+23.76\%} & \textbf{+48} \\
    w/o Tools & Claude-Sonnet-4.5 & 71/81 & 56.90\% & 196 & 227 & +15.82\% & +31 \\
    \midrule
    w/o Forge & GPT-5-Mini & 34/81 & 17.17\% & 23 & 23 & +0.00\% & +0 \\
    \textbf{w/o Slither} & \textbf{GPT-5-Mini} & \textbf{67/81} & \textbf{53.37\%} & \textbf{80} & \textbf{101} & \textbf{+26.25\%} & \textbf{+21} \\
    w/o Tools & GPT-5-Mini & 69/81 & 55.89\% & 100 & 99 & -1.00\% & -1 \\
    \midrule
    w/o Forge & GPT-5.1 & 38/81 & 18.77\% & 23 & 22 & -4.35\% & -1 \\
    \textbf{w/o Slither} & \textbf{GPT-5.1} & \textbf{67/81} & \textbf{46.55\%} & \textbf{89} & \textbf{120} & \textbf{+34.83\%} & \textbf{+31} \\
    w/o Tools & GPT-5.1 & 61/81 & 31.65\% & 59 & 52 & -11.86\% & -7 \\
    \bottomrule
    \end{tabular}%
  }
  \vspace{-1ex}
\end{table}

\noindent
\textbf{Results.}
The results for functional correctness and security are shown in Table~\ref{tab:rq2-ablation} and Table~\ref{tab:rq2-vuln}, respectively.

\noindent
\underline{\textbf{Impact of Forge Feedback.}}
Removing Forge feedback leads to a dramatic drop in functional correctness across all models.
For instance, with Claude-Sonnet-4.5, Pass@1 drops from \textbf{64.39\%} to \textbf{26.18\%}.
Similar significant drops are observed for GPT-5-Mini (56.65\% $\rightarrow$ 17.17\%) and GPT-5.1 (54.71\% $\rightarrow$ 18.86\%).
This confirms that iterative refinement based on compiler and test feedback is the most critical component for generating correct smart contracts.

\noindent
\underline{\textbf{Impact of Slither Feedback.}}
Removing Slither feedback generally has a minor impact on functional correctness but significantly affects security.
In the standard comparison (Table~\ref{tab:rq2-vuln}), removing Slither increases vulnerabilities for Claude (+15.68\%) and GPT-5.1 (+4.10\%), though it shows a slight decrease for GPT-5-Mini (-6.40\%).
However, when we analyze the \textbf{Min-Vuln Round} (Table~\ref{tab:rq2-min-vuln}), the benefit of Slither becomes clear and consistent across all models.
Without Slither, the minimum achievable vulnerabilities increase significantly: +23.76\% for Claude, +26.25\% for GPT-5-Mini, and +34.83\% for GPT-5.1.
This demonstrates that Slither feedback effectively expands the solution space to include more secure variants, allowing the agent to find solutions with significantly fewer vulnerabilities.

\noindent
\underline{\textbf{Impact of File System Tools.}}
Removing file system tools results in a noticeable decline in functional correctness.
Pass@1 drops for all models (e.g., 64.39\% $\rightarrow$ 57.66\% for Claude, 54.71\% $\rightarrow$ 31.73\% for GPT-5.1).
This suggests that the ability to explore the project structure and read dependency files is important for generating code that integrates correctly with the existing codebase.

\intuition{
\textbf{Answer to RQ-2:}
(1) \textbf{Forge feedback} is the most critical component for functional correctness, with its removal causing a drastic drop in Pass@1 (e.g., >38\% drop for Claude).
(2) \textbf{Slither feedback} is essential for security. While its impact varies in the standard round, it consistently enables the generation of significantly more secure code in the Min-Vuln round (reducing vulnerabilities by 23\%-34\%).
(3) \textbf{File system tools} contribute to functional correctness by providing necessary context for dependency integration and for generating code that integrates seamlessly into existing projects.
}
\subsection{RQ-3 Distillation Effectiveness}
\label{sec:rq3}

\noindent
\textbf{Objective.}
We investigate whether the capabilities of our multi-agent system can be distilled into a smaller, more efficient model. Specifically, we fine-tune Qwen3-8B on the high-quality interaction trajectories generated by \toolname (using Claude-Sonnet-4.5, GPT-5-Mini, and GPT-5.1) and evaluate its performance against the pre-distillation base model and a larger baseline (Qwen3-32B).

\noindent
\textbf{Experimental Design.}
We perform full-parameter fine-tuning on the Qwen3-8B model using two datasets derived from \toolname's trajectories, as detailed in Section~\ref{sec:ft-dataset}.
The training was conducted on a server with 8 $\times$ Huawei Ascend 910B2 NPUs (64GB). To handle context limits efficiently under limited memory, we implemented two tokenization strategies for each dataset:
\begin{itemize}[leftmargin=*]
    \item \textbf{v1 (Forward Truncation)}: Retains the first 4K tokens after tokenization.
    \item \textbf{v2 (Backward Truncation)}: Based on v1, this version retains the last 4K tokens. This strategy is designed to ensure the model learns from the final output (the generated code), which is critical for distillation.
\end{itemize}
This resulted in four distilled models: {\toolname-tracker} (v1/v2) and {\toolname-mix} (v1/v2). 
We utilized a cluster with 10 $\times$ NVIDIA V100 GPUs (32GB). We deployed the models using vLLM and enabled a 128K context window via YaRN RoPE scaling to support extensive code generation tasks.

\noindent
\textbf{Results.}
Table~\ref{tab:rq3-distill} presents the comparison between the pre-distillation models and the post-distillation \toolname variants, evaluated on the held-out test set.

\begin{table}[htbp]
  \vspace{-1ex}
  \centering
  \caption{Distillation Effectiveness: Pre- vs. Post-Distillation Comparison}
  \label{tab:rq3-distill}
  \vspace{-1.3ex}
  \resizebox{.95\linewidth}{!}{
    \begin{tabular}{lccccc}
    \toprule
    \textbf{Model} & \textbf{Compile Rate} & \textbf{$\Delta$Compile} & \textbf{Pass@1 (\%)} & \textbf{$\Delta$Pass@1} & \textbf{Pass@1 (Mean$\pm$Std)} \\
    \midrule
    Qwen3-8B & 5.88\% & - & 0.33\% & - & 0.5000 $\pm$ 0.0000 \\
    Qwen3-32B & 35.29\% & - & 1.31\% & - & 0.3333 $\pm$ 0.3727 \\
    \midrule
    \toolname-tracker-v1 & 11.76\% & +5.88\% & 0.98\% & +0.66\% & 0.7500 $\pm$ 0.2500 \\
    \rowcolor{lightgray} \toolname-tracker-v2 & 17.65\% & +11.76\% & 1.31\% & +0.98\% & 0.6667 $\pm$ 0.2357 \\
    \toolname-mix-v1 & 11.76\% & +5.88\% & 0.66\% & +0.33\% & 0.5000 $\pm$ 0.5000 \\
    \toolname-mix-v2 & 11.76\% & +5.88\% & 0.98\% & +0.66\% & 0.7500 $\pm$ 0.2500 \\
    \bottomrule
    \end{tabular}%
  }
  \par\vspace{0.5ex}
  \footnotesize
  \raggedright
  \textit{Note:} $\Delta$ represents the improvement relative to the teacher model (Qwen3-8B Base).
  \vspace{-1ex}
\end{table}

We observe several key findings:
(1) \textbf{Significant Improvement over Base Model}: All distilled models significantly outperform the pre-trained Qwen3-8B. The base 8B model struggles with the task, achieving only 5.88\% compilation rate and 0.33\% Pass@1. In contrast, \toolname-tracker-v2 triples the compilation rate to 17.65\% and quadruples the Pass@1 to 1.31\%.
(2) \textbf{Matching Larger Models}: Remarkably, the best distilled 8B model (\toolname-tracker-v2) achieves a Pass@1 of 1.31\%, which exactly matches the performance of the much larger Qwen3-32B model. This demonstrates that distilling high-quality agent trajectories can effectively bridge the performance gap between smaller and larger models.
(3) \textbf{Effectiveness of Backward Truncation (v2)}: The v2 models consistently outperform their v1 counterparts (e.g., tracker-v2's 1.31\% vs tracker-v1's 0.98\%). This confirms our hypothesis that retaining the latter part of the context (which contains the agent's final code output) is crucial for effective learning.

\intuition{
\textbf{Answer to RQ-3:}
Distilling \toolname's high-quality interaction trajectories into a smaller model (Qwen3-8B) significantly enhances its code generation capabilities. The distilled 8B model matches the functional correctness (Pass@1) of a 4x larger model (Qwen3-32B), demonstrating the efficiency of our distillation approach.
This offers a fast path for scenarios requiring lower latency or deployment cost.
The result also suggests that the agentic approach can benefit smaller, open-source models.
}

\section{Related Work}
\label{sec:related_work}

\subsection{LLM-based Code Generation and Agents}
Large Language Models (LLMs) have evolved from solving competitive programming tasks~\cite{Li2022Competitionlevel,yin2025learning} to achieving state-of-the-art performance in general code generation~\cite{Nijkamp2023CodeGen,Guo2022UniXcoder,Peng2026RepoGenesis}.
Rigorous evaluations~\cite{Liu2023Your,Kang2023Large} highlight this progress, including ``System 2'' reasoning for program repair~\cite{Yin2024ThinkRepair,yang2025input}.
In smart contracts, efforts focus on fine-tuning and preference alignment~\cite{Peng2025PrefGen}.
However, single-pass generation often suffers from hallucinations~\cite{Zhang2025LLM} and compilation failures, despite inference optimization techniques~\cite{Guo2024When}. 
To address these limitations, agentic frameworks have emerged.
Approaches like MetaGPT~\cite{Hong2024MetaGPT} and ChatDev~\cite{Qian2024ChatDev} simulate development teams, while specialized agents handle repository navigation (AutoCodeRover~\cite{Zhang2024AutoCodeRover}), interface design (SWE-agent~\cite{Yang2024SWEagent}), and environment interaction (OpenHands~\cite{Wang2025OpenHands}).
Agents have also been applied to microservice generation~\cite{Peng2026RepoGenesis}, GUI testing~\cite{Liu2024Make}, code review~\cite{Wang2024Unity}, and fault localization~\cite{Qin2025SoapFL}.
Despite these advancements, a ``domain mismatch'' persists in smart contract development due to the lack of specialized verification loops.
\toolname bridges this gap by integrating domain-specific tools like Forge and Slither directly into the agent's feedback loop, i.e., ``Tool-Augmented Refinement''.

\subsection{Smart Contract Security and Vulnerability Detection}
Ensuring the security of smart contracts is paramount due to the immutable nature of blockchain and the financial assets at stake~\cite{peng2025mulchainenablingadvancedcrossmodal}. 
Traditional approaches rely on static analysis (e.g., Slither~\cite{Feist2019Slither,TrailofBits2025Slither}, Mythril~\cite{ConsensysDiligence2025Mythril}) and fuzzing (e.g., Foundry/Forge~\cite{FoundryContributors2025Foundry}, Echidna~\cite{Grieco2020Echidna}) to detect vulnerabilities. 
While powerful, effective utilization of these tools usually requires significant manual effort to interpret comprehensive reports and expert knowledge to remediate identified issues. 
Recently, researchers have explored using LLMs for vulnerability detection and repair. 
Chen et al.~\cite{Chen2025When} conducted an empirical study evaluating ChatGPT's performance on smart contract vulnerability detection, finding that while it achieves high recall, its precision is limited and robustness needs improvement.
Sun et al. proposed GPTScan~\cite{Sun2024GPTScan}, which combines GPT with static analysis to detect logic vulnerabilities.
Similarly, AdvScanner~\cite{Wu2024AdvSCanner} generates adversarial exploits, while SmartInv~\cite{Wang2024SmartInv} utilizes multimodal learning to infer security invariants.
Chen et al. introduced NumScout~\cite{Chen2025NumScout}, which employs LLM-pruning symbolic execution to unveil numerical defects in smart contracts, achieving 89.7\% precision.
ContractTinker~\cite{Wang2024ContractTinker} further explores LLM-empowered vulnerability repair.
However, these works primarily focus on detecting bugs in existing code. 
Complementary to vulnerability detection, automated test generation is crucial for ensuring functional correctness. Zhang et al. proposed Solmigrator~\cite{Zhang2025Automated}, which extracts test cases from on-chain contract usage and migrates them to newly developed contracts, achieving 96.3\% precision and 93.6\% accuracy.
\toolname integrates these security tools during the generation process. 
By treating tool outputs as feedback signals, our agents can proactively identify and fix vulnerabilities before the code is finalized, effectively shifting security to the left in the development lifecycle.


\section{Conclusion and Future Work}
\label{sec:conclusion}

In this paper, we introduced \toolname, a novel tool-augmented multi-agent framework for automated smart contract generation that addresses the critical gap between LLMs' generative capabilities and the rigorous correctness and security requirements of blockchain development.
Our extensive evaluation on the SolEval+ Benchmark demonstrates that \toolname significantly outperforms state-of-the-art LLMs (including GPT-5 and Claude-Sonnet-4.5) and existing agent frameworks. It achieves a remarkable Pass@1 rate of 64.39\% while reducing security vulnerabilities by nearly 39.77\% compared to human-written baselines. The ablation study confirms the critical role of our dual-loop refinement mechanism and context-aware tool usage in achieving these results. Furthermore, we showed that the high-quality interaction trajectories generated by \toolname can be effectively distilled into smaller, open-source models, paving the way for more accessible and efficient automated smart contract development. In future work, we plan to extend \toolname to support more complex, multi-contract systems and explore the integration of formal verification tools to provide even stronger security guarantees. We also aim to investigate the application of our tool-augmented agentic approach to other safety-critical domains beyond blockchain.

\section*{Data Availability}
The replication of this paper is publicly available~\cite{2026ReplicationRelease}.

\bibliographystyle{ACM-Reference-Format}
\bibliography{main}

@misc{2026ReplicationRelease,
  title = {Replication},
  year = 2026,
  url = {https://github.com/openpaperz/SolAgent}
}

@article{yin2025learning,
  title={Learning to align human code preferences},
  author={Yin, Xin and Ni, Chao and Yang, Xiaohu},
  journal={arXiv preprint arXiv:2507.20109},
  year={2025}
}

@misc{peng2025mulchainenablingadvancedcrossmodal,
  title={MulChain: Enabling Advanced Cross-Modal Queries in Hybrid-Storage Blockchains}, 
  author={Zhiyuan Peng and Xin Yin and Gang Wang and Chenhao Ying and Wei Chen and Xikun Jiang and Yibin Xu and Yuan Luo},
  year={2025},
  eprint={2502.18258},
  archivePrefix={arXiv},
  primaryClass={cs.DB},
  url={https://arxiv.org/abs/2502.18258}, 
}

@article{yang2025input,
  title={Input reduction enhanced llm-based program repair},
  author={Yang, Boyang and Ren, Luyao and Yin, Xin and Ren, Jiadong and Tian, Haoye and Jin, Shunfu},
  journal={arXiv preprint arXiv:2507.15251},
  year={2025}
}

@article{Azimi2025Systematic,
  title = {A Systematic Review on Smart Contracts Security Design Patterns},
  author = {Azimi, Sadaf and Golzari, Ali and Ivaki, Naghmeh and Laranjeiro, Nuno},
  year = 2025,
  journal = {Empirical Software Engineering},
  volume = {30},
  number = {4},
  pages = {95}
}

@misc{BinanceNews2025Cetus,
  title = {Cetus {{Protocol Halts Operations After}} \$260 {{Million Hack}}, {{Sending Sui Tokens Into Freefall}}},
  author = {Binance News},
  year = 2025,
  url = {https://www.binance.com/en/square/post/24590524479882}
}

@inproceedings{Chen2017Underoptimized,
  title = {Under-Optimized Smart Contracts Devour Your Money},
  booktitle = {2017 {{IEEE}} 24th {{International Conference}} on {{Software Analysis}}, {{Evolution}} and {{Reengineering}} ({{SANER}})},
  author = {Chen, Ting and Li, Xiaoqi and Luo, Xiapu and Zhang, Xiaosong},
  year = 2017,
  pages = {442--446}
}

@misc{Chen2021Evaluating,
  title = {Evaluating {{Large Language Models Trained}} on {{Code}}},
  author = {Chen, Mark and Tworek, Jerry and Jun, Heewoo and Yuan, Qiming and Pinto, Henrique Ponde de Oliveira and Kaplan, Jared and Edwards, Harri and Burda, Yuri and Joseph, Nicholas and Brockman, Greg and Ray, Alex and Puri, Raul and Krueger, Gretchen and Petrov, Michael and Khlaaf, Heidy and Sastry, Girish and Mishkin, Pamela and Chan, Brooke and Gray, Scott and Ryder, Nick and Pavlov, Mikhail and Power, Alethea and Kaiser, Lukasz and Bavarian, Mohammad and Winter, Clemens and Tillet, Philippe and Such, Felipe Petroski and Cummings, Dave and Plappert, Matthias and Chantzis, Fotios and Barnes, Elizabeth and {Herbert-Voss}, Ariel and Guss, William Hebgen and Nichol, Alex and Paino, Alex and Tezak, Nikolas and Tang, Jie and Babuschkin, Igor and Balaji, Suchir and Jain, Shantanu and Saunders, William and Hesse, Christopher and Carr, Andrew N. and Leike, Jan and Achiam, Josh and Misra, Vedant and Morikawa, Evan and Radford, Alec and Knight, Matthew and Brundage, Miles and Murati, Mira and Mayer, Katie and Welinder, Peter and McGrew, Bob and Amodei, Dario and McCandlish, Sam and Sutskever, Ilya and Zaremba, Wojciech},
  year = 2021,
  number = {arXiv:2107.03374},
  eprint = {2107.03374},
  primaryclass = {cs},
  archiveprefix = {arXiv},
  note = {arXiv:2107.03374 [cs]}
}

@article{Chen2024Angels,
  title = {Angels or Demons: {{Investigating}} and Detecting Decentralized Financial Traps on Ethereum Smart Contracts},
  author = {Chen, Jiachi and Hu, Jiang and Xia, Xin and Lo, David and Grundy, John and Gao, Zhipeng and Chen, Ting},
  year = 2024,
  journal = {Automated Software Engineering},
  volume = {31},
  number = {2},
  pages = {63}
}

@article{Chen2025NumScout,
  title = {{{NumScout}}: {{Unveiling}} Numerical Defects in Smart Contracts Using {{LLM-pruning}} Symbolic Execution},
  author = {Chen, Jiachi and Shao, Zhenzhe and Yang, Shuo and Shen, Yiming and Wang, Yanlin and Chen, Ting and Shan, Zhenyu and Zheng, Zibin},
  year = 2025,
  journal = {IEEE Transactions on Software Engineering},
  volume = {51},
  number = {5},
  pages = {1538--1553}
}

@article{Chen2025When,
  title = {When {{ChatGPT}} Meets Smart Contract Vulnerability Detection: {{How}} Far Are We?},
  author = {Chen, Chong and Su, Jianzhong and Chen, Jiachi and Wang, Yanlin and Bi, Tingting and Yu, Jianxing and Wang, Yanli and Lin, Xingwei and Chen, Ting and Zheng, Zibin},
  year = 2025,
  journal = {ACM Transactions on Software Engineering and Methodology},
  volume = {34},
  number = {4},
  pages = {100:1--100:30}
}

@misc{ConsensysDiligence2025Mythril,
  title = {Mythril},
  author = {Consensys Diligence},
  year = 2025,
  url = {https://github.com/ConsenSys/mythril}
}

@inproceedings{Daspe2024Benchmarking,
  title = {Benchmarking Large Language Models for Ethereum Smart Contract Development},
  booktitle = {2024 6th {{Conference}} on {{Blockchain Research}} \& {{Applications}} for {{Innovative Networks}} and {{Services}} ({{BRAINS}})},
  author = {Daspe, Etienne and Durand, Mathis and Hatin, Julien and Bradai, Salma},
  year = 2024,
  pages = {1--4}
}

@inproceedings{Feist2019Slither,
  title = {Slither: {{A}} Static Analysis Framework for Smart Contracts},
  booktitle = {2019 {{IEEE}}/{{ACM}} 2nd {{International Workshop}} on {{Emerging Trends}} in {{Software Engineering}} for {{Blockchain}} ({{WETSEB}})},
  author = {Feist, Josselin and Grieco, Gustavo and Groce, Alex},
  year = 2019,
  pages = {8--15}
}

@misc{FoundryContributors2025Foundry,
  title = {Foundry: {{A}} Blazing Fast, Portable and Modular Toolkit for {{Ethereum}} Application Development},
  author = {Foundry Contributors},
  year = 2025,
  url = {https://github.com/foundry-rs/foundry}
}

@misc{Github2025Github,
  title = {Github {{Copilot}}},
  author = {Github},
  year = 2025,
  url = {https://github.com/features/copilot}
}

@inproceedings{Grieco2020Echidna,
  title = {Echidna: Effective, Usable, and Fast Fuzzing for Smart Contracts},
  booktitle = {Proceedings of the 29th {{ACM SIGSOFT International Symposium}} on {{Software Testing}} and {{Analysis}}},
  author = {Grieco, Gustavo and Song, Will and Cygan, Artur and Feist, Josselin and Groce, Alex},
  year = 2020,
  pages = {557--560}
}

@inproceedings{Guo2022UniXcoder,
  title = {{{UniXcoder}}: {{Unified}} Cross-Modal Pre-Training for Code Representation},
  booktitle = {Proceedings of the 60th {{Annual Meeting}} of the {{Association}} for {{Computational Linguistics}} ({{Volume}} 1: {{Long Papers}})},
  author = {Guo, Daya and Lu, Shuai and Duan, Nan and Wang, Yanlin and Zhou, Ming and Yin, Jian},
  year = 2022,
  pages = {7212--7225}
}

@inproceedings{Guo2024When,
  title = {When to Stop? {{Towards}} Efficient Code Generation in {{LLMs}} with Excess Token Prevention},
  booktitle = {Proceedings of the 33rd {{ACM SIGSOFT International Symposium}} on {{Software Testing}} and {{Analysis}}},
  author = {Guo, Lianghong and Wang, Yanlin and Shi, Ensheng and Zhong, Wanjun and Zhang, Hongyu and Chen, Jiachi and Zhang, Ruikai and Ma, Yuchi and Zheng, Zibin},
  year = 2024,
  pages = {1073--1085}
}

@inproceedings{Hong2024MetaGPT,
  title = {{{MetaGPT}}: {{Meta}} Programming for a Multi-Agent Collaborative Framework},
  booktitle = {The {{Twelfth International Conference}} on {{Learning Representations}}},
  author = {Hong, Sirui and Zhuge, Mingchen and Chen, Jonathan and Zheng, Xiawu and Cheng, Yuheng and Wang, Jinlin and Zhang, Ceyao and Wang, Zili and Yau, Steven Ka Shing and Lin, Zijuan and Zhou, Liyang and Ran, Chenyu and Xiao, Lingfeng and Wu, Chenglin and Schmidhuber, J{\"u}rgen},
  year = 2024
}

@inproceedings{Huang2024Large,
  title = {Large Language Models Cannot Self-Correct Reasoning Yet},
  booktitle = {The {{Twelfth International Conference}} on {{Learning Representations}}},
  author = {Huang, Jie and Chen, Xinyun and Mishra, Swaroop and Zheng, Huaixiu Steven and Yu, Adams Wei and Song, Xinying and Zhou, Denny},
  year = 2024
}

@inproceedings{Inala2022Faultaware,
  title = {Fault-Aware Neural Code Rankers},
  booktitle = {Proceedings of the 36th {{International Conference}} on {{Neural Information Processing Systems}}},
  author = {Inala, Jeevana Priya and Wang, Chenglong and Yang, Mei and Codas, Andres and Encarnaci{\'o}n, Mark and Lahiri, Shuvendu K and Musuvathi, Madanlal and Gao, Jianfeng},
  year = 2022,
  pages = {13419--13432}
}

@inproceedings{Kang2023Large,
  title = {Large Language Models Are Few-Shot Testers: {{Exploring LLM-based}} General Bug Reproduction},
  booktitle = {Proceedings of the 45th {{International Conference}} on {{Software Engineering}}},
  author = {Kang, Sungmin and Yoon, Juyeon and Yoo, Shin},
  year = 2023,
  pages = {2312--2323}
}

@article{Li2022Competitionlevel,
  title = {Competition-Level Code Generation with {{AlphaCode}}},
  author = {Li, Yujia and Choi, David and Chung, Junyoung and Kushman, Nate and Schrittwieser, Julian and Leblond, R{\'e}mi and Eccles, Tom and Keeling, James and Gimeno, Felix and Dal Lago, Agustin and Hubert, Thomas and Choy, Peter and {de Masson d'Autume}, Cyprien and Babuschkin, Igor and Chen, Xinyun and Huang, Po-Sen and Welbl, Johannes and Gowal, Sven and Cherepanov, Alexey and Molloy, James and Mankowitz, Daniel J. and Sutherland Robson, Esme and Kohli, Pushmeet and {de Freitas}, Nando and Kavukcuoglu, Koray and Vinyals, Oriol},
  year = 2022,
  journal = {Science},
  volume = {378},
  number = {6624},
  pages = {1092--1097}
}

@misc{Li2023ModelScopeagent,
  title = {{{ModelScope-agent}}: {{Building}} Your Customizable Agent System with Open-Source Large Language Models},
  author = {Li, Chenliang and Chen, Hehong and Yan, Ming and Shen, Weizhou and Xu, Haiyang and Wu, Zhikai and Zhang, Zhicheng and Zhou, Wenmeng and Chen, Yingda and Cheng, Chen and Shi, Hongzhu and Zhang, Ji and Huang, Fei and Zhou, Jingren},
  year = 2023,
  number = {arXiv:2309.00986},
  eprint = {2309.00986},
  primaryclass = {cs},
  archiveprefix = {arXiv},
  note = {arXiv:2309.00986 [cs]}
}

@misc{Li2025DeepCode,
  title = {{{DeepCode}}: {{Open}} Agentic Coding},
  author = {Li, Zongwei and Li, Zhonghang and Guo, Zirui and Ren, Xubin and Huang, Chao},
  year = 2025,
  number = {arXiv:2512.07921},
  eprint = {2512.07921},
  primaryclass = {cs},
  archiveprefix = {arXiv},
  note = {arXiv:2512.07921 [cs]}
}

@article{Liao2024MathbfA^3A3CodGen,
  title = {\textbackslash{{mathbfA}}\textasciicircum{{3A3-CodGen}}: {{A}} Repository-Level Code Generation Framework for Code Reuse with Local-Aware, Global-Aware, and Third-Party-Library-Aware},
  author = {Liao, Dianshu and Pan, Shidong and Sun, Xiaoyu and Ren, Xiaoxue and Huang, Qing and Xing, Zhenchang and Jin, Huan and Li, Qinying},
  year = 2024,
  journal = {IEEE Transactions on Software Engineering},
  volume = {50},
  number = {12},
  pages = {3369--3384}
}

@inproceedings{Liu2023Your,
  title = {Is Your Code Generated by {{ChatGPT}} Really Correct? {{Rigorous}} Evaluation of Large Language Models for Code Generation},
  booktitle = {Advances in Neural Information Processing Systems},
  author = {Liu, Jiawei and Xia, Chunqiu Steven and Wang, Yuyao and ZHANG, {\relax LINGMING}},
  year = 2023,
  volume = {36},
  pages = {21558--21572}
}

@inproceedings{Liu2024Make,
  title = {Make {{LLM}} a Testing Expert: {{Bringing}} Human-like Interaction to Mobile {{GUI}} Testing via Functionality-Aware Decisions},
  booktitle = {Proceedings of the {{IEEE}}/{{ACM}} 46th {{International Conference}} on {{Software Engineering}}},
  author = {Liu, Zhe and Chen, Chunyang and Wang, Junjie and Chen, Mengzhuo and Wu, Boyu and Che, Xing and Wang, Dandan and Wang, Qing},
  year = 2024,
  pages = {1--13}
}

@inproceedings{Nijkamp2023CodeGen,
  title = {{{CodeGen}}: {{An}} Open Large Language Model for Code with Multi-Turn Program Synthesis},
  booktitle = {The {{Eleventh International Conference}} on {{Learning Representations}}},
  author = {Nijkamp, Erik and Pang, Bo and Hayashi, Hiroaki and Tu, Lifu and Wang, Huan and Zhou, Yingbo and Savarese, Silvio and Xiong, Caiming},
  year = 2023
}

@inproceedings{Peng2025PrefGen,
  title = {{{PrefGen}}: {{A}} Preference-Driven Methodology for Secure yet Gas-Efficient Smart Contract Generation},
  booktitle = {{{IEEE}}/{{ACM International Conference}} on {{Automated Software Engineering}}},
  author = {Peng, Zhiyuan and Yin, Xin and Zhou, Zijie and Ying, Chenhao and Ni, Chao and Luo, Yuan},
  year = 2025
}

@inproceedings{Peng2025SolEval,
  title = {{{SolEval}}: {{Benchmarking}} Large Language Models for Repository-Level Solidity Smart Contract Generation},
  booktitle = {Proceedings of the 2025 {{Conference}} on {{Empirical Methods}} in {{Natural Language Processing}}},
  author = {Peng, Zhiyuan and Yin, Xin and Qian, Rui and Lin, Peiqin and Liu, YongKang and Zhang, Hao and Ying, Chenhao and Luo, Yuan},
  year = 2025,
  pages = {4388--4411}
}

@misc{Peng2026RepoGenesis,
  title = {{{RepoGenesis}}: {{Benchmarking}} End-to-End Microservice Generation from Readme to Repository},
  author = {Peng, Zhiyuan and Yin, Xin and Zhao, Pu and Yang, Fangkai and Wang, Lu and Jia, Ran and Chen, Xu and Lin, Qingwei and Rajmohan, Saravan and Zhang, Dongmei},
  year = 2026,
  number = {arXiv:2601.13943},
  eprint = {2601.13943},
  primaryclass = {cs},
  archiveprefix = {arXiv},
  note = {arXiv:2601.13943 [cs]}
}

@inproceedings{Perez2021Smart,
  title = {Smart Contract Vulnerabilities: {{Vulnerable}} Does Not Imply Exploited},
  booktitle = {30th {{USENIX Security Symposium}} ({{USENIX Security}} 21)},
  author = {Perez, Daniel and Livshits, Benjamin},
  year = 2021,
  pages = {1325--1341}
}

@inproceedings{Qian2024ChatDev,
  title = {{{ChatDev}}: {{Communicative}} Agents for Software Development},
  booktitle = {Proceedings of the 62nd {{Annual Meeting}} of the {{Association}} for {{Computational Linguistics}} ({{Volume}} 1: {{Long Papers}})},
  author = {Qian, Chen and Liu, Wei and Liu, Hongzhang and Chen, Nuo and Dang, Yufan and Li, Jiahao and Yang, Cheng and Chen, Weize and Su, Yusheng and Cong, Xin and Xu, Juyuan and Li, Dahai and Liu, Zhiyuan and Sun, Maosong},
  year = 2024,
  pages = {15174--15186}
}

@article{Qin2025SoapFL,
  title = {{{SoapFL}}: {{A}} Standard Operating Procedure for {{LLM-based}} Method-Level Fault Localization},
  author = {Qin, Yihao and Wang, Shangwen and Lou, Yiling and Dong, Jinhao and Wang, Kaixin and Li, Xiaoling and Mao, Xiaoguang},
  year = 2025,
  journal = {IEEE Transactions on Software Engineering},
  volume = {51},
  number = {4},
  pages = {1173--1187}
}

@misc{QwenTeam2025QwenAgent,
  title = {Qwen-{{Agent}}: {{A Framework}} for {{Agents}} Based on {{Qwen}}},
  author = {Qwen Team},
  year = 2025,
  url = {https://github.com/QwenLM/Qwen-Agent}
}

@inproceedings{Sharma2023Mixedmethods,
  title = {A Mixed-Methods Study of Security Practices of Smart Contract Developers},
  booktitle = {32nd {{USENIX Security Symposium}} ({{USENIX Security}} 23)},
  author = {Sharma, Tanusree and Zhou, Zhixuan and Miller, Andrew and Wang, Yang},
  year = 2023,
  pages = {2545--2562}
}

@inproceedings{Sun2024GPTScan,
  title = {{{GPTScan}}: {{Detecting}} Logic Vulnerabilities in Smart Contracts by Combining {{GPT}} with Program Analysis},
  booktitle = {Proceedings of the {{IEEE}}/{{ACM}} 46th {{International Conference}} on {{Software Engineering}}},
  author = {Sun, Yuqiang and Wu, Daoyuan and Xue, Yue and Liu, Han and Wang, Haijun and Xu, Zhengzi and Xie, Xiaofei and Liu, Yang},
  year = 2024,
  pages = {1--13}
}

@misc{TrailofBits2025Slither,
  title = {Slither: {{Static Analyzer}} for {{Solidity}}},
  author = {{Trail of Bits}},
  year = 2025,
  url = {https://github.com/crytic/slither}
}

@inproceedings{Wang2024ContractTinker,
  title = {{{ContractTinker}}: {{LLM-empowered}} Vulnerability Repair for Real-World Smart Contracts},
  booktitle = {Proceedings of the 39th {{IEEE}}/{{ACM International Conference}} on {{Automated Software Engineering}}},
  author = {Wang, Che and Zhang, Jiashuo and Gao, Jianbo and Xia, Libin and Guan, Zhi and Chen, Zhong},
  year = 2024,
  pages = {2350--2353}
}

@inproceedings{Wang2024SmartInv,
  title = {{{SmartInv}}: {{Multimodal}} Learning for Smart Contract Invariant Inference},
  booktitle = {2024 {{IEEE Symposium}} on {{Security}} and {{Privacy}} ({{SP}})},
  author = {Wang, Sally Junsong and Pei, Kexin and Yang, Junfeng},
  year = 2024,
  pages = {2217--2235}
}

@inproceedings{Wang2024Unity,
  title = {Unity Is Strength: {{Collaborative LLM-based}} Agents for Code Reviewer Recommendation},
  booktitle = {Proceedings of the 39th {{IEEE}}/{{ACM International Conference}} on {{Automated Software Engineering}}},
  author = {Wang, Luqiao and Zhou, Yangtao and Zhuang, Huiying and Li, Qingshan and Cui, Di and Zhao, Yutong and Wang, Lu},
  year = 2024,
  pages = {2235--2239}
}

@inproceedings{Wang2025OpenHands,
  title = {{{OpenHands}}: {{An}} Open Platform for {{AI}} Software Developers as Generalist Agents},
  booktitle = {The {{Thirteenth International Conference}} on {{Learning Representations}}},
  author = {Wang, Xingyao and Li, Boxuan and Song, Yufan and Xu, Frank F. and Tang, Xiangru and Zhuge, Mingchen and Pan, Jiayi and Song, Yueqi and Li, Bowen and Singh, Jaskirat and Tran, Hoang H. and Li, Fuqiang and Ma, Ren and Zheng, Mingzhang and Qian, Bill and Shao, Yanjun and Muennighoff, Niklas and Zhang, Yizhe and Hui, Binyuan and Lin, Junyang and Brennan, Robert and Peng, Hao and Ji, Heng and Neubig, Graham},
  year = 2025
}

@article{Wang2025Unity,
  title = {Unity Is Strength: {{Enhancing}} Precision in Reentrancy Vulnerability Detection of Smart Contract Analysis Tools},
  author = {Wang, Zexu and Chen, Jiachi and Zheng, Peilin and Zhang, Yu and Zhang, Weizhe and Zheng, Zibin},
  year = 2025,
  journal = {IEEE Transactions on Software Engineering},
  volume = {51},
  number = {1},
  pages = {1--13}
}

@inproceedings{Wu2024AdvSCanner,
  title = {{{AdvSCanner}}: {{Generating}} Adversarial Smart Contracts to Exploit Reentrancy Vulnerabilities Using {{LLM}} and Static Analysis},
  booktitle = {Proceedings of the 39th {{IEEE}}/{{ACM International Conference}} on {{Automated Software Engineering}}},
  author = {Wu, Yin and Xie, Xiaofei and Peng, Chenyang and Liu, Dijun and Wu, Hao and Fan, Ming and Liu, Ting and Wang, Haijun},
  year = 2024,
  pages = {1019--1031}
}

@article{Yang2024SWEagent,
  title = {{{SWE-agent}}: {{Agent-computer}} Interfaces Enable Automated Software Engineering},
  author = {Yang, John and Jimenez, Carlos E. and Wettig, Alexander and Lieret, Kilian and Yao, Shunyu and Narasimhan, Karthik and Press, Ofir},
  year = 2024,
  journal = {Advances in Neural Information Processing Systems},
  volume = {37},
  pages = {50528--50652}
}

@inproceedings{Yin2024ThinkRepair,
  title = {{{ThinkRepair}}: {{Self-directed}} Automated Program Repair},
  booktitle = {Proceedings of the 33rd {{ACM SIGSOFT International Symposium}} on {{Software Testing}} and {{Analysis}}},
  author = {Yin, Xin and Ni, Chao and Wang, Shaohua and Li, Zhenhao and Zeng, Limin and Yang, Xiaohu},
  year = 2024,
  pages = {1274--1286}
}

@inproceedings{Yu2024CoderEval,
  title = {{{CoderEval}}: {{A}} Benchmark of Pragmatic Code Generation with Generative Pre-Trained Models},
  booktitle = {Proceedings of the {{IEEE}}/{{ACM}} 46th {{International Conference}} on {{Software Engineering}}},
  author = {Yu, Hao and Shen, Bo and Ran, Dezhi and Zhang, Jiaxin and Zhang, Qi and Ma, Yuchi and Liang, Guangtai and Li, Ying and Wang, Qianxiang and Xie, Tao},
  year = 2024,
  pages = {1--12}
}

@inproceedings{Zhang2023Demystifying,
  title = {Demystifying Exploitable Bugs in Smart Contracts},
  booktitle = {2023 {{IEEE}}/{{ACM}} 45th {{International Conference}} on {{Software Engineering}} ({{ICSE}})},
  author = {Zhang, Zhuo and Zhang, Brian and Xu, Wen and Lin, Zhiqiang},
  year = 2023,
  pages = {615--627}
}

@inproceedings{Zhang2024AutoCodeRover,
  title = {{{AutoCodeRover}}: {{Autonomous}} Program Improvement},
  booktitle = {Proceedings of the 33rd {{ACM SIGSOFT International Symposium}} on {{Software Testing}} and {{Analysis}}},
  author = {Zhang, Yuntong and Ruan, Haifeng and Fan, Zhiyu and Roychoudhury, Abhik},
  year = 2024,
  pages = {1592--1604}
}

@inproceedings{Zhang2025Automated,
  title = {Automated Test Generation for Smart Contracts via On-Chain Test Case Augmentation and Migration},
  booktitle = {2025 {{IEEE}}/{{ACM}} 47th {{International Conference}} on {{Software Engineering}} ({{ICSE}})},
  author = {Zhang, Jiashuo and Chen, Jiachi and Grundy, John and Gao, Jianbo and Wang, Yanlin and Chen, Ting and Guan, Zhi and Chen, Zhong},
  year = 2025,
  pages = {1947--1959}
}

@article{Zhang2025LLM,
  title = {{{LLM}} Hallucinations in Practical Code Generation: {{Phenomena}}, Mechanism, and Mitigation},
  author = {Zhang, Ziyao and Wang, Chong and Wang, Yanlin and Shi, Ensheng and Ma, Yuchi and Zhong, Wanjun and Chen, Jiachi and Mao, Mingzhi and Zheng, Zibin},
  year = 2025,
  journal = {Proc. ACM Softw. Eng.},
  volume = {2},
  number = {ISSTA},
  pages = {ISSTA022:481--ISSTA022:503}
}

@article{Zhao2025SWIFT,
  title = {{{SWIFT}}: {{A}} Scalable Lightweight Infrastructure for Fine-Tuning},
  author = {Zhao, Yuze and Huang, Jintao and Hu, Jinghan and Wang, Xingjun and Mao, Yunlin and Zhang, Daoze and Jiang, Zeyinzi and Wu, Zhikai and Ai, Baole and Wang, Ang and Zhou, Wenmeng and Chen, Yingda},
  year = 2025,
  journal = {Proceedings of the AAAI Conference on Artificial Intelligence},
  volume = {39},
  number = {28},
  pages = {29733--29735}
}

\clearpage
\newpage
\setcounter{page}{1}


\end{document}